\documentclass{emulateapj}
\usepackage{color}
\usepackage{rotating}
\usepackage{gensymb}
\usepackage[mathscr]{euscript}

 \newcommand{\vecv}{\mbox{\boldmath $v$} {}}
\newcommand{\vecV}{\mbox{\boldmath $V$} {}}
\newcommand{\vecr}{\mbox{\boldmath $r$} {}}

\newcommand{\vece}{\mbox{\boldmath $e$} {}}
\newcommand{\vecF}{\mbox{\boldmath $F$} {}}

\newcommand{\qmfive}{\mbox{$q_{\scriptscriptstyle -5}$} {}}

\usepackage{epsfig}
\usepackage{amsmath}
\newif\ifAMStwofonts
\shorttitle{Inspirals in retrograde orbits}
\shortauthors{S\'anchez-Salcedo}
\begin{document}

\title{Orbital evolution of gas-driven inspirals with extreme mass-ratios:\\ retrograde eccentric orbits}

\author{
  F. J. S\'anchez-Salcedo\altaffilmark{1} 
 }

\altaffiltext{1}{Instituto de Astronom\'{\i}a, Universidad Nacional Aut\'onoma de M\'exico, A. P. 70-264,
Mexico City 04510, Mexico \email{(jsanchez@astro.unam.mx)}}

\begin{abstract}
Using two-dimensional simulations, we compute the torque and rate of work (power) on a low-mass 
gravitational body, with softening length $R_{\rm soft}$, embedded in a gaseous disk when its orbit is 
eccentric and retrograde with respect to the disk. We explore orbital eccentricities $e$ between
$0$ and $0.6$.
We find that the power has its maximum at $e\simeq 0.25(h/0.05)^{2/3}$, where $h$ is the aspect ratio of the disk.
We show that the power and the torque converge to the values predicted in the local (non-resonant)
approximation of the dynamical friction (DF) when $R_{\rm soft}$ tends to zero. 
For retrograde inspirals with mass ratios $\lesssim 5\times 10^{-4}$ embedded
in disks with $h\geq 0.025$, our simulations suggest that (i) the rate of inspiral barely depends
on the orbital eccentricity and (ii) the local approximation provides the value of this inspiral
rate within a factor of $1.5$. The implications of the results for the
orbital evolution of extreme mass-ratio inspirals are discussed.

\end{abstract}


\section{Introduction}
\label{sec:intro}

At the center of galaxies, stars can draine into the central supermassive black hole (SMBH) 
due to two-body diffusion, resonant relaxation and dynamical friction (DF) with the surrounding 
material (mainly dark matter and gas) \cite[e.g.][]{hop06}. 
Compact objects (COs), such as stellar remnants and stellar mass black holes (stellar BHs) 
can inspiral into a SMBH and emit gravitational waves, which could be detected by the Laser 
Interferometer Space Antenna (LISA) \cite[e.g.,][]{fin00,ama17}.

In the presence of accretion disks as those in active galactic nuclei (AGN), 
stars and COs can experience gravitational torques that can accelerate the radial migration 
towards the center \cite[e.g.,][]{arm02,koc11}.
COs may belong to the nuclear cluster \citep{mck11} 
or may have formed inside the AGN star-forming disk \cite[e.g.,][]{lev07}.

Nuclear cluster COs may have prograde as well as retrograde orbits with respect to the AGN accretion 
disk. COs born in the star-forming disk are expected to move on prograde orbits.
Still, gravitational scattering between them or with other objects (including a SMBH binary companion) 
may excite large orbital eccentricities  \cite[e.g.,][]{pap01,bre19}.
In principle, it is plausible that some of the estimated $10^{3}$ BHs
of mass $(7-10)M_{\odot}$ that resides within $0.1$ pc of the central BH may be scattered to 
retrograde eccentric orbits and can even counter-rotate with respect to the accretion disk.

Disk COs may counter-rotate with respect to the AGN accretion disk if the AGN disk is rejuvenated
with captured gas clouds having uncorrelated angular momentum \citep{ima18,imp19},
as occurs at galactic scales in some galaxies  
\cite[e.g.,][]{gar03,cor14,mar18}.

COs and intermediate mass BHs can also rotate with high inclinations, if they
are brought to the galactic center anchored in an inclined stellar cluster \citep{por03,por06,kim03,gur05,
ant12,ant14,arc18}.
After the stellar cluster is destroyed by
tidal forces, all the COs and intermediate mass BHs residing in the stellar cluster will be
spread out in inclined orbits.

The evolution of the semi-major axis $a$, the eccentricity $e$ and the inclination $i$ of
a perturber due to the tidal interaction with the disk has been studied intensively in the
context of protoplanetary disks  \cite[e.g.,][]{art93,pap00,gol03,tan04,cre07,mar09,bit10,bit11,bit13}.
For $e$ or $i$ larger than the aspect ratio of the disk $h\equiv H/R$ (where $H$ is the scaleheight
of the disk at radius $R$), the perturber moves supersonically with respect to the gas.  For that
reason, a DF approach has been invoked to describe the interaction between
the disk and a body in inclined or eccentric orbits \cite[e.g.,][]{pap02,mut11,rei12,ama16}.
In \citet{san19}, we find that a simple model based on DF describes
the orbital evolution of bodies in coplanar ($i=0$) eccentric orbits ($h<e\lesssim 0.6$), provided
that the ratio between the mass of the perturber and the mass of the central object (denoted by $q$)
is sufficiently small. For typical protoplanetary disks, this occurs for planets with $q\lesssim 10^{-4}$.

For highly-inclined circular orbits,
\citet{rei12} considers the aerodynamical and gravitational drag forces on a planet when it
crosses the protoplanetary disk. For orbits with $i=45^{\circ}, 90^{\circ}$ and $155^{\circ}$,
he finds good agreement between the gravitational drag force measured in numerical simulations and the 
force predicted using a formula based on DF arguments. \citet{xia13}
carry out a set of numerical simulations of the orbital evolution
of a gravitational perturber in a circular orbit, for the full range of inclinations. 
They argue that the qualitative behaviour of the results can be interpreted using simple formula based
on DF.   

The limiting case $i=180^{\circ}$ (retrograde orbit) and $e=0$ (circular orbit), was studied in
\citet{iva15} and \citet{san18}.
In this case, the perturber moves supersonically (Mach numbers of $\simeq 40-100$). As a result, 
the perturber catches its own wake repeatedly and, in fact, the pull imparted by
the wake ahead of the perturber cannot be ignored unless the mass ratio $q$ is small enough
\citep{san18}.

In the general case (arbitrary values of $i$ and $e$, but larger than $h$), one expects that if $q$ is small enough, 
most of the contribution to the drag force arises from the portion of the wake just at the rear of the 
body and, therefore, the DF approximation should be valid to quantify the components of the 
drag force and thereby the evolution of $a$, $e$ and $i$. 

In order to complete our picture on the applicability and limitations of an approach based on DF,
which is impulsive and non-resonant,
we use numerical simulations to evaluate the components of the drag force when the 
orbit is retrograde ($i=180^{\circ}$) and eccentric. Interestingly, for certain disk parameters typical for AGN
disks, the DF formula predicts that the eccentricity may grow. 
This stands in sharp contrast to the rapid eccentricity damping seen in the prograde case.
We wish to quantify to what extend the predictions based on DF considerations are reliable.

The structure of the paper is as follows. 
In Section \ref{sec:basic}, we describe our system and provide the basic equations. 
In Section \ref{sec:LA}, we present the DF framework in its local approximation (hereafter LA) and 
make some predictions. A comparison between predictions and the results of hydrodynamical simulations 
are given in Section \ref{sec:simulations}. The implications for the evolution of COs embedded in
AGN disks are discussed in Section \ref{sec:compactcase}. Finally, we summarize our conclusions 
in Section \ref{sec:conclusions}.

\begin{figure}
\includegraphics[width=92mm,height=80mm]{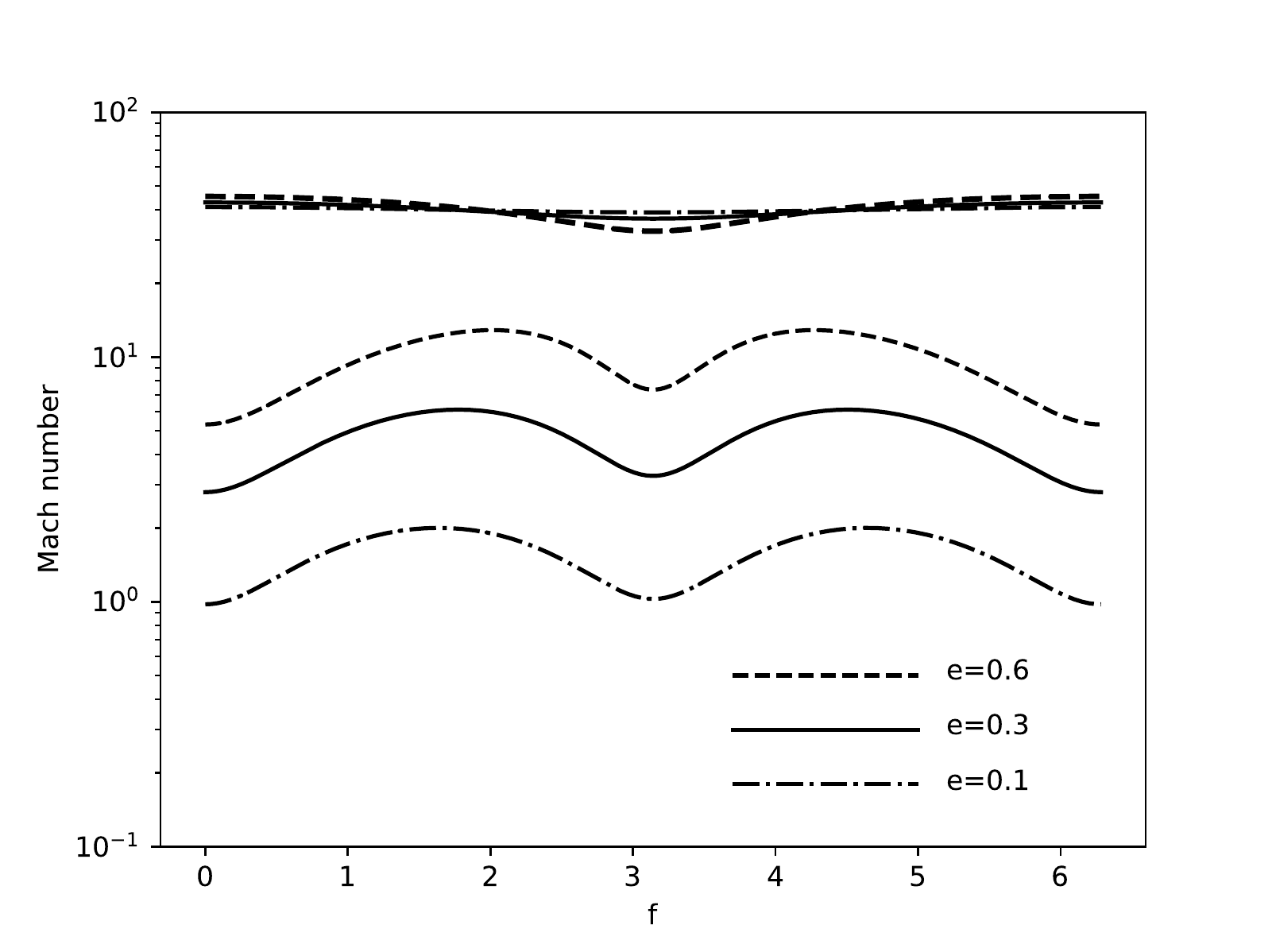}
  \caption{Local Mach number ${\mathcal{M}}$ versus true anomaly $f$ along the
Keplerian orbit described by a body in the midplane of a locally isothermal disk with 
aspect ratio $h=0.05$, for prograde (lower curves) and retrograde motion (upper 
curves).
 }
\vskip 0.25cm
\label{fig:Mach_f}
\end{figure}

\section{Description of the model: Basic equations}
\label{sec:basic}

Our system consists of an accretion disk around a central SMBH with
mass $M_{\bullet}\simeq 10^{5}-10^{7}$ M$_{\odot}$, plus a CO (e.g., a stellar BH), the perturber, with mass 
$M_{p}\simeq 1-10$ M$_{\odot}$. Therefore,
the mass ratio $\qmfive\equiv q/10^{-5}$ is between $0.01$ to $10$. These systems are referred to
as extreme mass ratio inspirals (EMRIs). The mass of the disk is assumed to be much 
smaller than $M_{\bullet}$. The orbital plane of the CO is taken coplanar with the disk.
The orbit can be prograde or retrograde.

The CO will exchange energy and angular momentum with the disk through the tidal interaction.
As a consequence, the semimajor axis $a$ and the eccentricity $e$ of the CO will change with time.
Let $P$ denote the power, i.e.~the energy change of the CO per unit of time, and $T$ the torque 
imparted on the CO. The evolution equations for $a$ and $e$ are given by
\begin{equation}
\overline{\frac{da}{dt}}=\frac{2\overline{P}}{a\omega^{2}M_{p}},
\end{equation}
and
\begin{equation}
\overline{\frac{de}{dt}}= \frac{\eta^{2}\overline{B}}{e a^{2} \omega^{2}M_{p}},
\end{equation}
where $\eta\equiv \sqrt{1-e^2}$, $\omega=\sqrt{GM_{\bullet}/a^{3}}$ and
$\overline{B}\equiv\overline{P}-\omega \eta^{-1} \overline{T}$ (e.g., Murray \& Dermott 1999).
The bar over a variable denotes orbit-averaged values.
In these equations, we have applied the sign convention that the torque is positive (negative) 
when the CO gains (loses) angular momentum.

\begin{figure*}
\includegraphics[width=199mm,height=75mm]{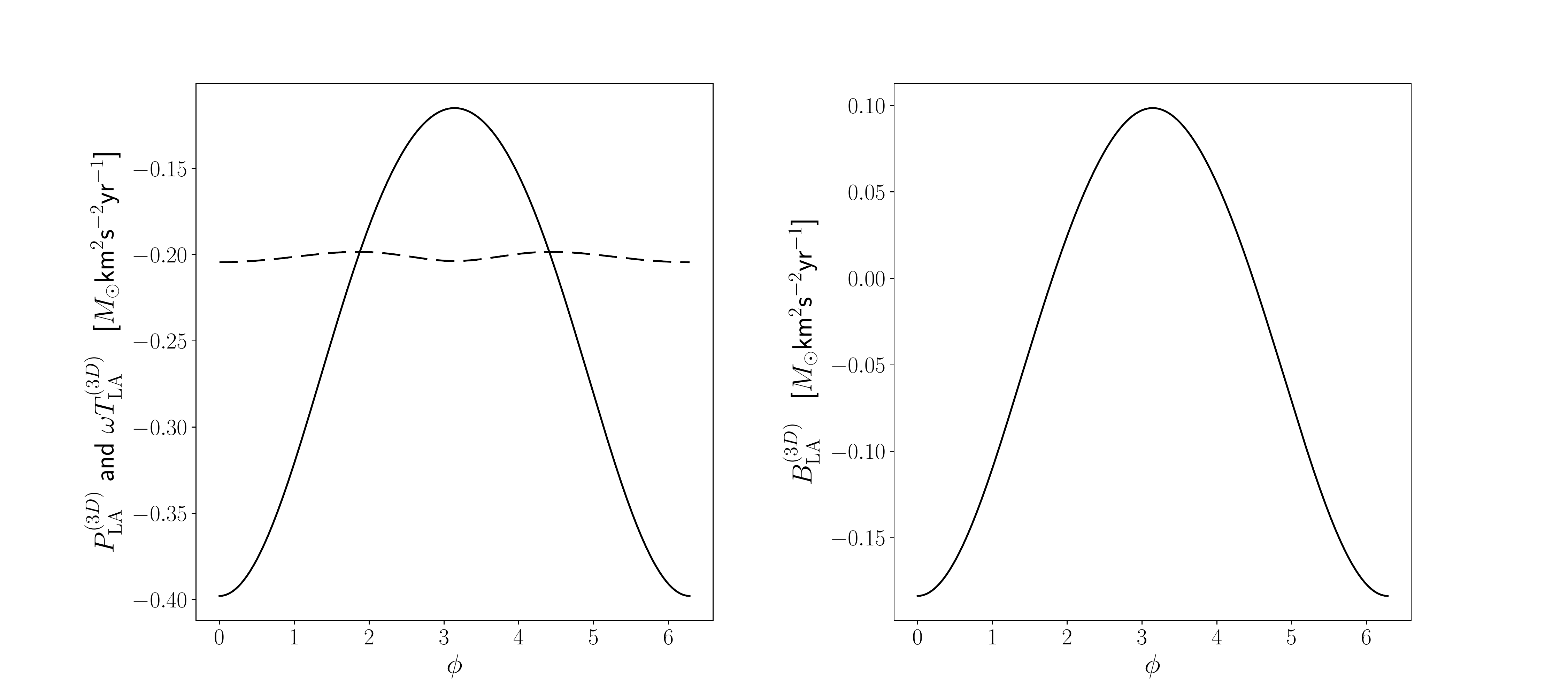}
  \caption{Predictions using the LA. The left panel shows the power (solid line) and 
the torque (dashed line), imparted on a CO with orbital eccentricity $0.3$, as a function of the orbital 
phase in a disk with $\alpha=3/2$ and $\lambda=1/2$. The right panel shows $B\equiv P-\omega \eta^{-1}T$. 
 }
\vskip 0.25cm
\label{fig:pw_tq_outer_disk}
\end{figure*}

\begin{figure*}
\includegraphics[width=199mm,height=80mm]{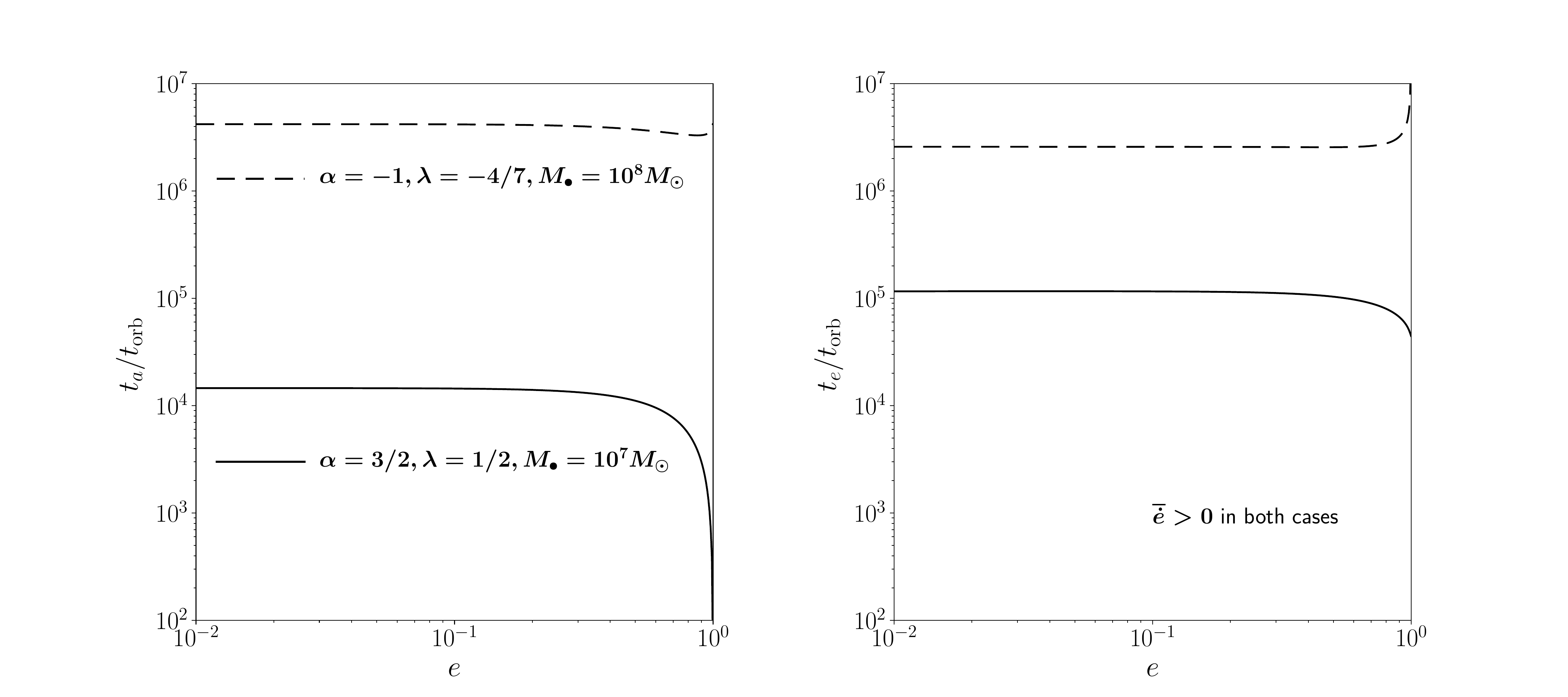}
  \caption{Migration timescale (left panel) and eccentricity growth timescale (right panel) as a function
of the orbital eccentricity for a $10M_{\odot}$ BH in the outer parts of an AGN disk with a central SMBH of
$10^{7}M_{\odot}$ (solid lines) and in the inner parts of an AGN disk with a central SMBH of $10^{8}M_{\odot}$
(dashed lines). 
 }
\vskip 0.25cm
\label{fig:timescales_inner_outer}
\end{figure*}

The migration timescale $t_{a}$, in units of the orbital period $t_{\rm orb}\equiv 2\pi/\omega$, is
\begin{equation}
\frac{t_{a}}{t_{\rm orb}}\equiv \frac{1}{t_{\rm orb}} \bigg|\frac{a}{\overline{\dot{a}}}\bigg| = \frac{a^{2}\omega^{3}M_{p}}{4\pi |\overline{P}|}.
\label{eq:taua}
\end{equation}
The orbital eccentricity changes on the timescale
\begin{equation}
\frac{t_{e}}{t_{\rm orb}}\equiv \frac{1}{t_{\rm orb}} \bigg|\frac{e}{\overline{\dot{e}}}\bigg| =
\frac{e^{2} a^{2}\omega^{3}M_{p}}{2\pi \eta^{2}|\overline{B}|}.
\label{eq:taue}
\end{equation} 
By their definitions, the timescales $t_{a}$ and $t_{e}$ are always positive. We anticipate that the
eccentricity may be damped or excited, depending on the disk parameters. Therefore,
we will give $t_{e}$ and specify the sign of $\overline{\dot{e}}$.

The response of the disk to the presence of the CO depends on the relative velocity 
between the CO and the disk. We define the Mach number $\mathcal{M}$
as the ratio $V_{\rm rel}/c_{s}$, where
$V_{\rm rel}$ is the velocity of the perturber relative to the local gas and $c_{s}$ the local 
sound speed. Figure \ref{fig:Mach_f} shows $\mathcal{M}$ as a function of the true 
anomaly $f$ (the pericenter is at $f=0$ and the apocenter is at $f=\pi$). 
We have assumed that the CO describes an elliptical orbit and the disk aspect ratio is constant 
($h=0.05$) so that the isothermal sound speed is $c_{s}=hR\Omega$, where
$\Omega$ is the Keplerian angular velocity $\Omega (R)=\sqrt{GM_{\bullet}/R^{3}}$. 
From Fig. \ref{fig:Mach_f}, we see that the motion
for retrograde orbits is always supersonic regardless the value of $e$.
A difference between prograde and retrograde rotation is the orbital
position where ${\mathcal{M}}$ achieves its maximum value. 
For retrograde orbits, the maximum of ${\mathcal{M}}$ occurs at pericenter,
whereas it occurs at $f\simeq 2$ and $4.5$ for prograde orbits.
Another difference is that the orbital average ${\mathcal{M}}$ increases with $e$ for
prograde orbits, whereas it is essentially independent of $e$ for retrograde orbits.

Given their low $q$ and high $\mathcal{M}$, EMRIs in retrograde orbits cannot open a gap in the disk 
\citep{mck14,iva15,san18}
and, in addition, their accretion radii $R_{\rm acc}$ are generally much smaller than 
the vertical scaleheight $H$ of the disk. For instance, consider a retrograde EMRI at a radial
distance $R_{p}$ embedded in a disk with constant $h$. The relative velocity of the CO with respect 
to the gas is $V_{\rm rel}^{2}\simeq 4GM_{\bullet}/R_{p}$ (being this expression more accurate for small values of $e$). Therefore, $R_{\rm acc}  \equiv 2GM_{p}/V_{\rm rel}^{2} \simeq q R_{p}/2$. 
In terms of $H$, we have $R_{\rm acc}/H\simeq q/(2h)$. For EMRIs with $\qmfive\leq 10$ and $h$
between $0.02$ and $0.05$, we obtain $R_{\rm acc}/H \lesssim 2.5\times 10^{-3}$.

\section{The local approximation in 3D disks}
\label{sec:LA}

In the local approximation (LA), we apply the DF formula at every point of the orbit,
ignoring the curvature of the spiral wave behind the body. 
In \citet{san18}, we have studied the range of validity of the LA
for perturbers in retrograde and circular orbit. On the other hand, the case of prograde and eccentric
orbits was presented in \citet{san19}. These studies demonstrate that the LA can
predict the power and the torque provided that $q$ is small enough.
We note that in the retrograde circular case, there are no Lindblad
resonances, but they appear when the orbit is eccentric \citep{iva15,nix15}.

\begin{figure}
\includegraphics[width=90mm,height=80mm]{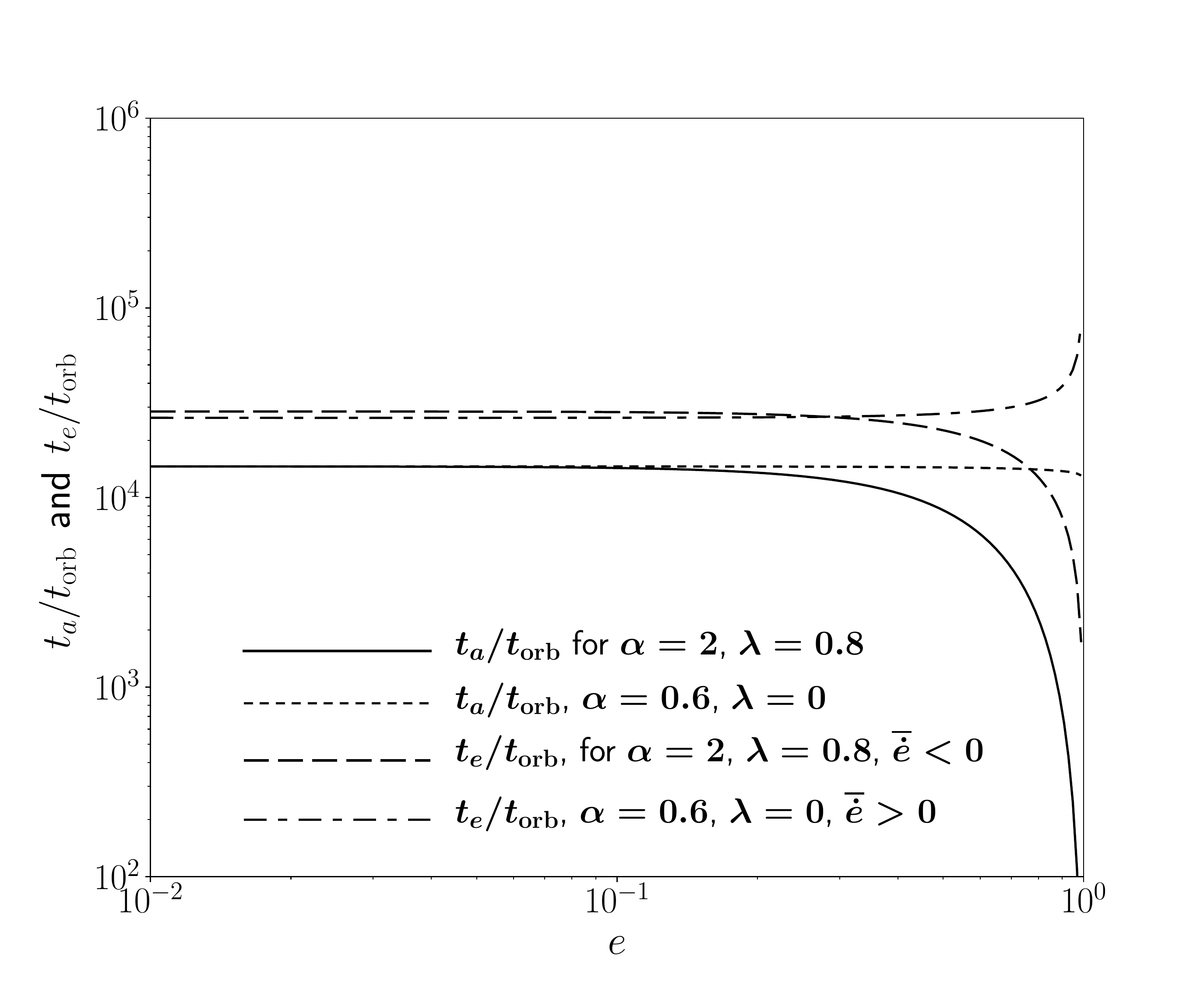}
  \caption{Evolution timescales $t_{a}$ and $t_{e}$ (in units of $t_{\rm orb}$)
for different combinations of $\alpha$ and $\lambda$, keeping the rest of the parameters fixed.}
\vskip 0.25cm
\label{fig:same_edot}
\end{figure}

If the LA were also valid for retrograde and eccentric orbits, then it would be easy
to find $t_{a}$ and $t_{e}$ as follows.
The force $\vecF_{\rm LA}^{(3D)}$ acting on a perfect accretor in the LA is
\begin{equation}
\vecF_{\rm LA}^{(3D)}=\frac{\sqrt{8\pi}\Sigma_{p} (GM_{p})^{2} \ln\Lambda_{p}}{V_{\rm rel}^{3}H_{p}} 
\, \vecV_{\rm rel},
\label{eq:FLA3D}
\end{equation}
where  the subscript $p$ indicates evaluation of the variable at the location of the perturber
\citep{can13,san18}. Here $\Sigma$ is the unperturbed
surface density of the disk, $H$ is its vertical scaleheight and $\Lambda=7.15H/R_{\rm acc}$.
In the derivation of Eq. (\ref{eq:FLA3D}), it was assumed 
that the disk volume density is $\rho(R,z)=\rho_{0}(R)\exp(-z^{2}/2H^{2})$.
The superscript $3D$ denotes that the 3D structure of the disk has been included.

We now assume that the equatorial plane of the disk is at $z=0$ and that it rotates counterclockwise
in a Keplerian fashion (we ignore deviations from the Keplerian rotation arising from the pressure
gradient). The unperturbed velocity of the gas is $\vecv_{g}=R\Omega \tilde{\vece}_{\phi}$ and 
the relative velocity is $\vecV_{\rm rel}=\vecv_{g}-\vecv_{p}$, where 
\begin{equation}
\vecv_{p}=\frac{p a\omega}{\eta}\left(e\sin\phi \hat{\vece}_{R} +
(1+e\cos\phi)\hat{\vece}_{\phi}\right),
\label{eq:vel_perturber}
\end{equation}
and $p=+1$ for prograde orbits and $p=-1$ for retrograde orbits (recall that
$\eta\equiv (1-e^{2})^{1/2}$). We have assumed that the 
pericenter is at $\phi=0$.
Although we are mainly interested in the retrograde case, we give
the expressions for both prograde and retrograde cases to highlight the differences.

Using Eqs. (\ref{eq:FLA3D}) and (\ref{eq:vel_perturber}), the power and the torque are given by
\begin{equation}
\begin{aligned}
P_{\rm LA}^{(3D)}& =\vecv_{p}\cdot \vecF_{\rm LA}^{(3D)}=
\frac{\sqrt{8\pi} p\eta q^{2} \omega^{3}a^{5}\Sigma_{p} \ln \Lambda_{p}} {H_{p}} \\
& \times \frac{-p e^{2}\sin^{2} \phi +\xi \hat{\xi}}{[e^{2}\sin^{2}\phi +\hat{\xi}^{2}]^{3/2}}
\end{aligned}
\label{eq:power3D}
\end{equation}
and
\begin{equation}
\begin{aligned}
T_{\rm LA}^{(3D)}& =p \hat{\vece}_{z}\cdot (\vecr_{p}\times \vecF_{\rm LA}^{(3D)})=
\frac{\sqrt{8\pi} p\eta^{4} q^{2} \omega^{2}a^{5}\Sigma_{p} \ln \Lambda_{p}} {H_{p}} \\
& \times \frac{\hat{\xi}}{\xi (e^{2}\sin^{2} \phi +\hat{\xi}^{2})^{3/2}},
\end{aligned}
\label{eq:torque3D}
\end{equation}
where $\xi(\phi) \equiv 1+e\cos \phi$ and $\hat{\xi}(\phi)=\sqrt{\xi}-p\xi$.
It is simple to show that the power and the torque are both negative at any orbital position if
$p=-1$.

In the following we consider some disk models that have been used to
describe protoplanetary disks and disks around the central BH in AGNs. These
models assume that the surface density and the scaleheight of the disk are given by
power laws. We suppose that 
\begin{equation}
\Sigma = \Sigma_{a}\left(\frac{R}{a}\right)^{-\alpha}
\end{equation}
and 
\begin{equation}
H = H_{a} \left(\frac{R}{a}\right)^{1+\lambda}
\end{equation}
where $\Sigma_{a}$ and $H_{a}$ are the surface density and the scaleheight of the disk at $R=a$,
respectively.

Simplified models of the structure of Keplerian viscous disks around SMBHs suggest $\alpha=3/2$
and $\lambda=1/2$ at distances $>10^{3} R_{\rm Sch}$, where $R_{\rm Sch}$ is the Schwarzschild radius 
of the central SMBH \cite[e.g.,][]{goo03,sir03}.
Figure \ref{fig:pw_tq_outer_disk} 
shows the predicted power and torque, in the LA, as a function of the orbital phase 
$\phi$ when the EMRI is retrograde and has $\qmfive=0.1$, $a=0.1$ pc and $e=0.3$. 
The remainder of the parameters are $M_{\bullet}=10^{7}M_{\odot}$,
$\Sigma_{a}=5\times 10^{6}M_{\odot}$pc$^{-2}$ and $H_{a}=1.4\times 10^{-3}$ pc.
According to Figure \ref{fig:pw_tq_outer_disk}, the
eccentricity is excited at apocenter, because the torque is more negative than the power ($B>0$).
On the contrary, the eccentricity decreses at pericenter ($B<0$). The orbital average 
$de/dt$ is positive (albeit very small: $\overline{B}=0.0046 M_{\odot}$ km$^{2}$ s$^{-2}$ yr$^{-1}$), 
implying that the eccentricity is excited.

For this model, we have computed how $t_{a}$ and $t_{e}$ depend on eccentricity 
(see Figure \ref{fig:timescales_inner_outer}). 
We find that $t_{a}$ and $t_{e}$ are almost constant between $e=0$ and $e=0.7$. 
We note that $t_{e}\sim 10 t_{a}$ for eccentricities in the range $0<e<0.8$.  Therefore, if the LA
is correct, we expect that, in the retrograde case, the migration takes place at 
almost constant eccentricity. This is in sharp contrast with the prograde case, where the orbit 
circularizes on a timescale short compared to the migration timescale (typically $t_{e}=0.1t_{a}$, 
e.g., Cresswell \& Nelson 2008).

At $R<10^{3} R_{\rm Sch}$, the models of \citet{sir03} predict $\alpha=-1$ and 
$\lambda=-4/7$. Being the surface density greater at apocenter, the positive value of $de/dt$ at apocenter 
is enhanced in the retrograde case. Figure \ref{fig:timescales_inner_outer} shows $t_{a}$ and $t_{e}$ 
in this part of the disk for $p=-1$,
$M_{\bullet}=10^{8}M_{\odot}$, $M_{p}=10M_{\odot}$, $a=0.005$ pc, $\Sigma_{a}=7\times 10^{8}M_{\odot}$pc$^{-2}$ and $H_{a}=8.5\times 10^{-5}$ pc. We find again 
that $\bar{\dot{e}}>0$ (the eccentricity grows) but now $t_{e}\simeq t_{a}$.

The eccentricity may be damped for certain combinations of $\alpha$ and $\lambda$, if they are
sufficiently large. Figure \ref{fig:same_edot} compares
the timescales for $\alpha=0.6$ and $\lambda=0$ with those for $\alpha=2$ and $\lambda=0.8$,
with the remainder of the parameters ($p, q, M_{\bullet}, a, \Sigma_{a}, H_{a}$)
being the same.  In the first case, $\overline{\dot{e}}$ is positive,
whereas it is negative for the second set of parameters, but both cases have approximately the same $t_{e}$
at $e<0.4$.

It is now clear that the LA provides a very useful framework to predict 
$t_{a}$ and $t_{e}$ in a rather simple way. It is therefore crucial to study its validity
domain.

\begin{figure}
\includegraphics[width=95mm,height=110mm]{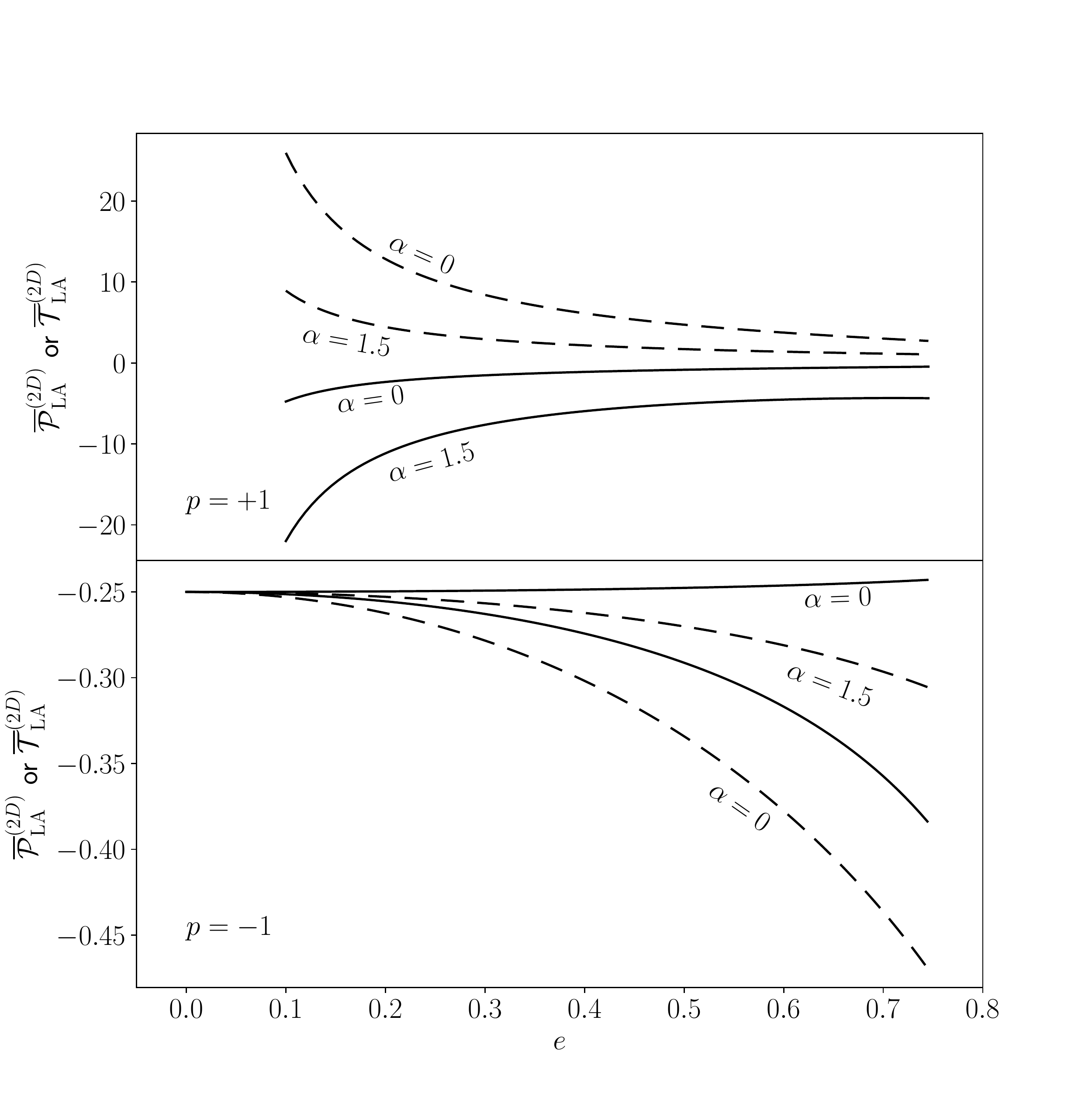}
  \caption{Orbit average power (solid lines) and torque (dashed lines)  as 
a function of the eccentricity in the LA, for prograde (top panel)
and retrograde (bottom panel) orbits. The disk has $h=$const and $\alpha$ was set to either $0$ or $1.5$. 
 }
\vskip 0.25cm
\label{fig:mean_pw_tq_LA}
\end{figure}

\section{Numerical experiments}
\label{sec:simulations}

The aim of this Section is to explore the conditions under which a local description can
be used to estimate the tidal forces exerted on a retrograde perturber. Since the LA essentially
ignores 2D effects, mainly the differential rotation and the curvature terms (the curvature of the wake 
and the curvature of perturber orbit),
it is sufficient to consider 2D disks. In fact, once the range of validity of the LA
is determined in 2D disks, the results can be extended to more realistic 3D disks.
This will be done in Section \ref{sec:compactcase}.

The response of the disk to the gravitational potential $\Phi_{p}$ of the perturber (the secondary) is simulated
using the code FARGO3D, which is a publicly available code\footnote{FARGO3D is available at http://fargo.in2p3.fr.} 
\citep{ben16}.
The perturber is placed on a fixed retrograde orbit with eccentricity $e$. 
We use polar coordinates $(R,\phi$), where $R$ is measured from the central object.

The potential $\Phi_{p}$ is modeled by introducing a softening length $R_{\rm soft}$:
\begin{equation}
\Phi_{p} = -\frac{GM_{p}}{\sqrt{(\vecr-\vecr_{p})^{2}+R_{\rm soft}^{2}}},
\end{equation}
where $\vecr_{p}$ is the position of the perturber.
Strictly, we are not simulating a point-mass particle as a BH, but just an extended non-accreting
body. Nevertheless, it is simple to extend the results to accreting point-mass objects 
(see \S \ref{sec:compactcase}).
For simplicity, we will take $R_{\rm soft}={\mathcal{E}} H$, where $H\equiv c_{s}/\Omega$ is the vertical
scaleheight of the disk and ${\mathcal{E}}$ is a constant. We will also assume that the aspect ratio
$h$ is constant over $R$; this condition fixes the radial profile of the temperature
of the disk. All together, $R_{\rm soft}\propto H\propto R$.

We will use dimensionless power ${\mathcal{P}}$ and torque ${\mathcal{T}}$ defined as
\begin{equation}
{\mathcal{P}}=\frac{{\mathcal{E}}h}{\pi q^{2}\omega^{3}a^{4} 
\Sigma_{a}} P,
\end{equation}
and
\begin{equation}
{\mathcal{T}}= \frac{{\mathcal{E}}h}{\pi q^{2}\omega^{2}a^{4} 
\Sigma_{a}} T.
\end{equation}

\begin{figure}
\includegraphics[width=92mm,height=70mm]{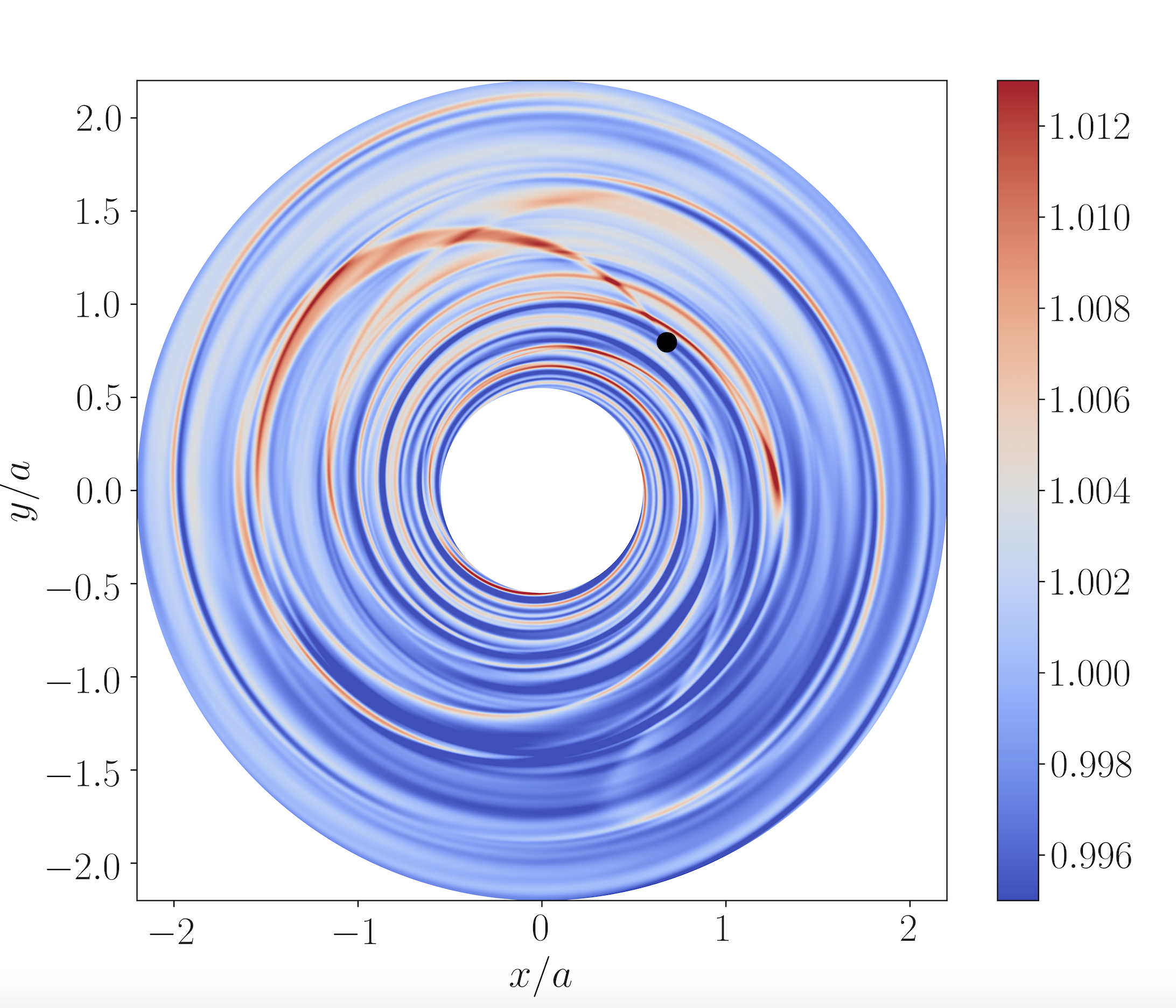}
  \caption{Color map of the density of the disk,
for a perturber with $\qmfive=1$, $e=0.6$ and
${\mathcal{E}}=0.3$. The black dot marks the position of the perturber, which is moving clockwise
from apocenter to pericenter. The computational domain is $0.12a<R<5.2a$, but the figure only
shows the ring $0.55a<R<2.2a$. At $R<0.55a$, the density spirals are far too thin for the {\it image}
resolution.
 }
\vskip 0.25cm
\label{fig:disk_map}
\end{figure}

In terms of dimensionless quantities, the timescales are
\begin{equation}
\frac{t_{a}}{t_{\rm orb}}= \frac{{\mathcal{E}} h}{4\pi q q_{d}} \frac{1}{|{\overline{\mathcal{P}}}|},
\end{equation}
and
\begin{equation}
\frac{t_{e}}{t_{\rm orb}}= \frac{e^{2}  {\mathcal{E}} h}{2\pi \eta^{2} q q_{d}} 
\frac{1}{|{\overline{\mathcal{B}}}|},
\label{eq:te_B}
\end{equation}
with $q_{d}\equiv \pi a^{2} \Sigma_{a}/M_{\bullet}$ and
${\mathcal{B}}\equiv  {\mathcal{P}}-\eta^{-1} {\mathcal{T}}$.

\begin{figure*}
\includegraphics[width=199mm,height=85mm]{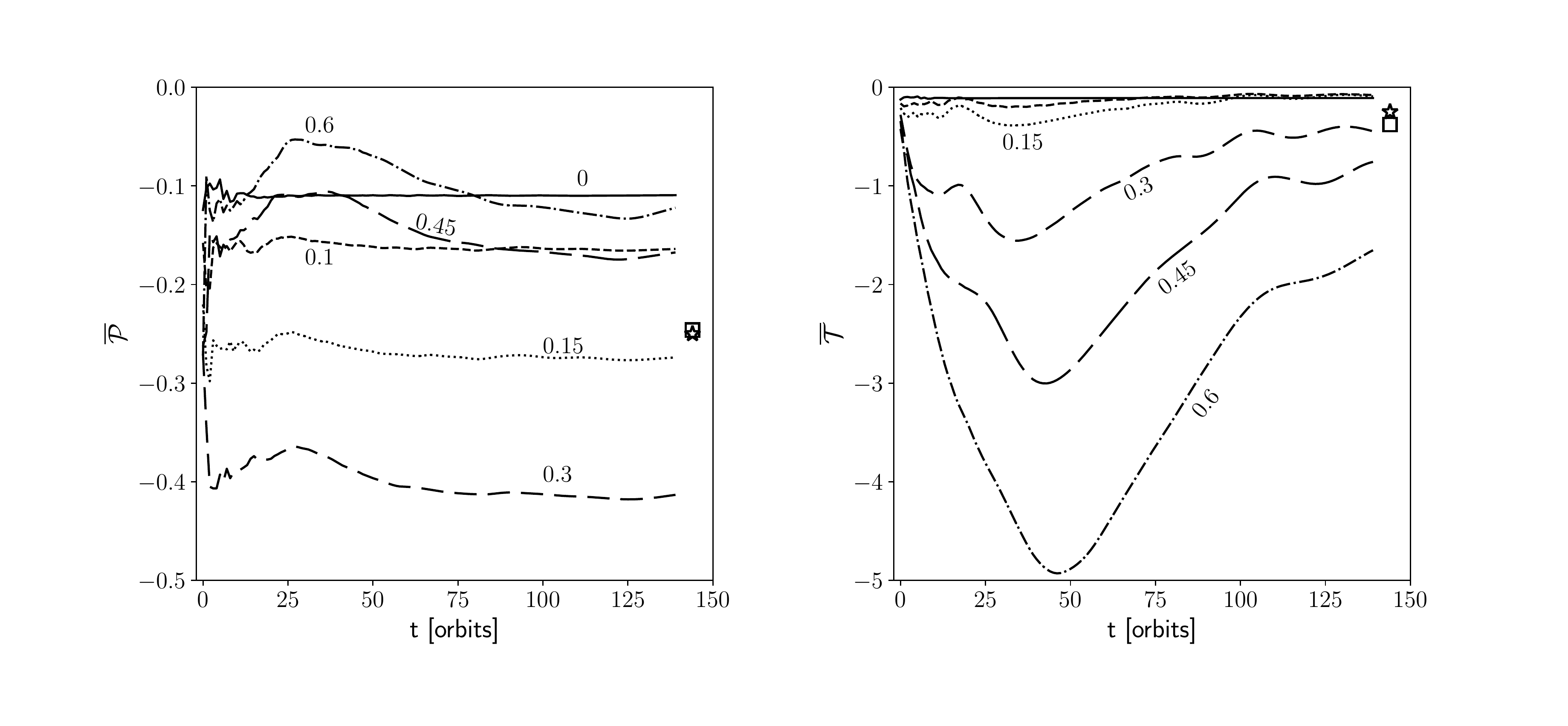}
  \caption{Evolution of the power (left panel) and torque (right panel) using $\alpha=0$, $h=0.05$, 
$\qmfive=1$ and ${\mathcal{E}}=0.6$. Different curves are for different eccentricities. The value
of the eccentricity is given at each curve. The symbols on the right side of each panel
indicate the values in the LA for $e=0$ (stars) and $e=0.6$ (squares). For intermediate
eccentricities, the LA values lie in between.
 }
\vskip 0.25cm
\label{fig:diff_ecc}
\end{figure*}

\begin{figure}
\hskip -0.30cm
\includegraphics[width=99mm,height=85mm]{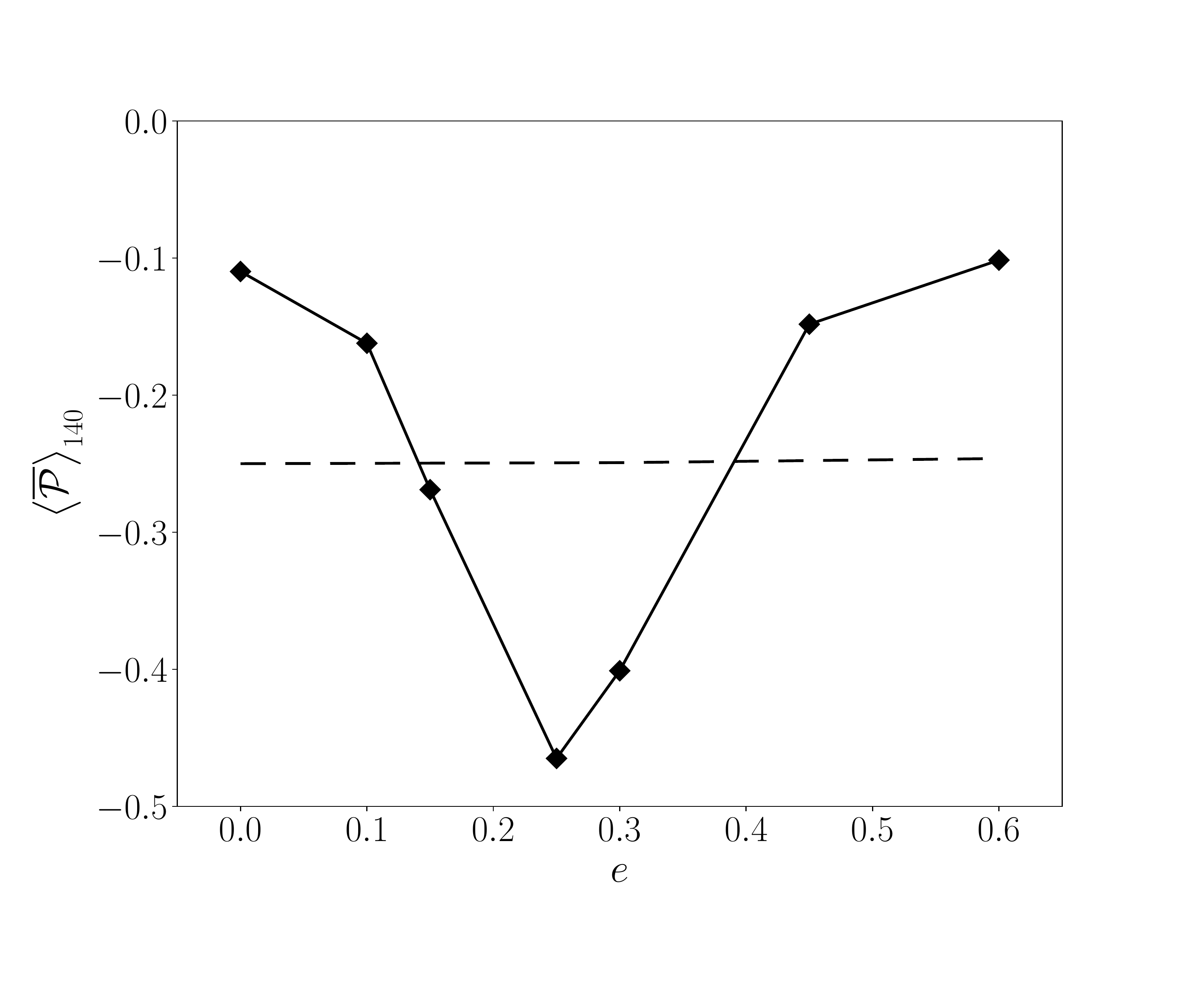}
\caption{Mean value of the power between the $5$th and $140$th orbits for the models
in Figure \ref{fig:diff_ecc}. The dashed line indicates the power in the LA.}
\vskip 0.2cm
\label{fig:pw140}
\end{figure}

In the razor-thin (2D) disk model, the LA predicts the following dimensionless power and
torque:
\begin{equation}
{\mathcal{P}}_{\rm LA}^{(2D)}= 
\frac{p\xi^{1+\alpha} (-p e^{2}\sin^{2} \phi +\xi \hat{\xi})}{\eta^{1+2\alpha} (e^{2}\sin^{2}\phi +\hat{\xi}^{2})^{3/2}},
\label{eq:power_DF_model}
\end{equation}
and
\begin{equation}
{\mathcal{T}}_{\rm LA}^{(2D)}= 
\frac{p\eta^{2(1-\alpha)}\xi^{\alpha}\hat{\xi}}{(e^{2}\sin^{2} \phi +\hat{\xi}^{2})^{3/2}}.
\label{eq:torque_DF_model}
\end{equation}
Here we have used that the drag force on a body travelling supersonically in a rectilinear orbit inside a 2D layer 
of surface density $\Sigma$ is
\begin{equation}
\vecF_{\rm LA}^{(2D)}=\frac{\pi \Sigma G^{2}M_{p}^{2}}{R_{\rm soft} V_{\rm rel}^{3}}  \vecV_{\rm rel}
\label{eq:muto_Fdf_thin}
\end{equation}
\citep{mut11}.

For illustration, Figure  \ref{fig:mean_pw_tq_LA} shows $\overline{\mathcal{P}}_{\rm LA}^{(2D)}$ and 
$\overline{\mathcal{T}}_{\rm LA}^{(2D)}$ as a function of $e$,  
for $p=+1$ (prograde) and $p=-1$ (retrograde). As expected, 
$|\overline{\mathcal{P}}_{\rm LA}^{(2D)}|$ and $|\overline{\mathcal{T}}_{\rm LA}^{(2D)}|$
are smaller in the retrograde case, especially at low eccentricities. It is remarkable that for $p=-1$ 
and $\alpha=0$, $\overline{\mathcal{P}}_{\rm LA}^{(2D)}$ is almost constant with $e$.

\begin{figure}
\hskip -0.30cm
\includegraphics[width=99mm,height=85mm]{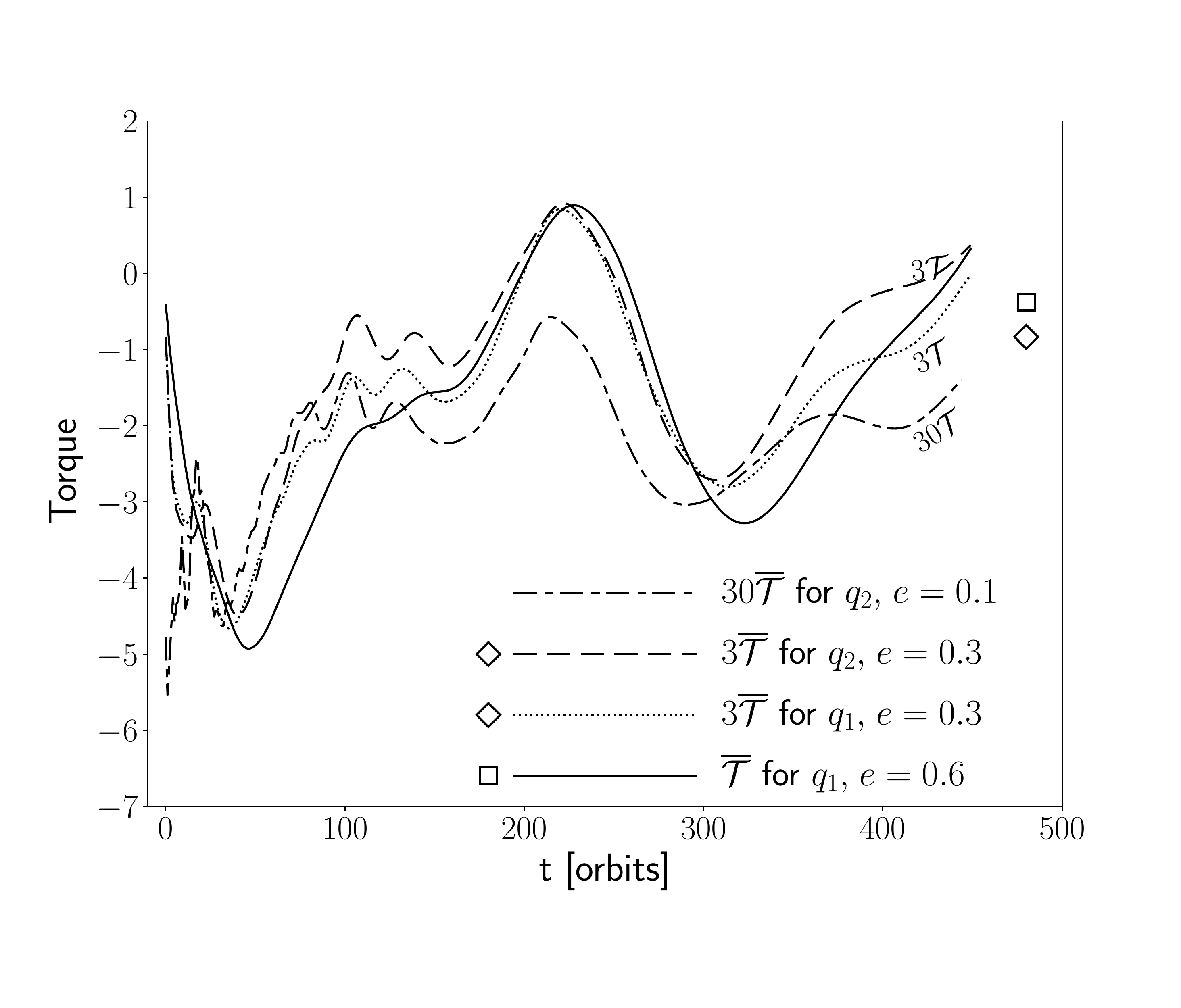}
\caption{Temporal evolution of the dimensionless torque for various combinations of $q$ and $e$. In all cases we take ${\mathcal{E}}=0.6$. 
The symbols at the right side indicate the corresponding values
in the LA. For $e=0.1$, the LA predicts $30\overline{\mathcal{T}}=-7.59$, but it is
not shown because it is outside the range.
 }
\vskip 0.2cm
\label{fig:tq_long_runs}
\end{figure}

\subsection{Range of parameters and other numerical issues}

We use values for $\qmfive$ between $1$ and $50$. Our reference value
for $h$ is $0.05$, but we explore other values in Section \ref{sec:diff_h}. We vary the eccentricity
between $0$ and $0.6$, and ${\mathcal{E}}$ between $0.06$ and $0.6$. For these parameters, 
the accretion radius of the perturber is $\lesssim 2.5\times 10^{-4}a$, which is
much smaller than $R_{\rm soft}={\mathcal{E}} h a=(3\times 10^{-3}-3\times 10^{-2})a$. 
In all our simulations we include a kinematic viscosity $\nu=10^{-5}\omega a^{2}$ constant 
through the disk.

The computational domain extends from $R_{\rm in}$ to $R_{\rm out}$. 
Appendix \ref{sec:BC} is devoted to assess the importance of the finite size of the domain
and to describe how the results depend on the boundary conditions.
Unless otherwise specified, we employ wave-killing zones at $R\in [R_{\rm in},1.3R_{\rm in}]$
and at $R\in [0.95R_{\rm out},R_{\rm out}]$, following the scheme described in de Val-Borro et al.
(2006).
Boundary effects are more pronounced for larger values of
${\mathcal{E}}$. When wave-damping boundary 
conditions are used, we find that $R_{\rm in}=0.2a$ and $R_{\rm out}=4a$ are adequate to 
compute the power and the torque within $100$ orbits even for ${\mathcal{E}}=0.6$
(see Appendix \ref{sec:BC}). 
In all the simulations presented in the remainder of the paper,
we use $R_{\rm out}=5.2a$, and we
take $R_{\rm in}=0.2a$ if $e\leq 0.3$, and $R_{\rm in}=0.12a$ if $e>0.3$.

In all the simulations, the perturber is inserted suddenly at $t=0$. 
In order to partially suppress transient effects during the relaxation process,
Appendix \ref{sec:adiab_perturber} contains the results of
simulations in which the mass of 
the perturber increases slowly over time until it reaches its final mass. 
In Appendix \ref{sec:adiab_perturber}, it is shown 
that those effects associated with relaxation are small.

We have studied the numerical convergence. For instance, for disks having $h=0.05$, 
we found that the measured power and torque do not change for 
$N_{\phi}\geq 2.5$ and $N_{r, \rm peri}\geq 2.5$, where $N_{\phi}$ and 
$N_{r,\rm peri}$ are 
the number of zones per $R_{\rm soft}$ in the azimuthal and radial directions, respectively.
We note that $N_{r,\rm peri}$ is computed at pericenter. In all the simulations presented in
this paper, both $N_{\phi}$ and $N_{r,\rm peri}$ are larger than $3$, typically $\sim 5$, 
to ensure that the resolution is adequate. We were especially careful to ensure that 
the resolution was enough to resolve the tightly-wound density perturbations
with small radial wavelength formed in the 
inner parts of the computational domain due to the strong Keplerian shear.

\begin{figure*}
\includegraphics[width=199mm,height=85mm]{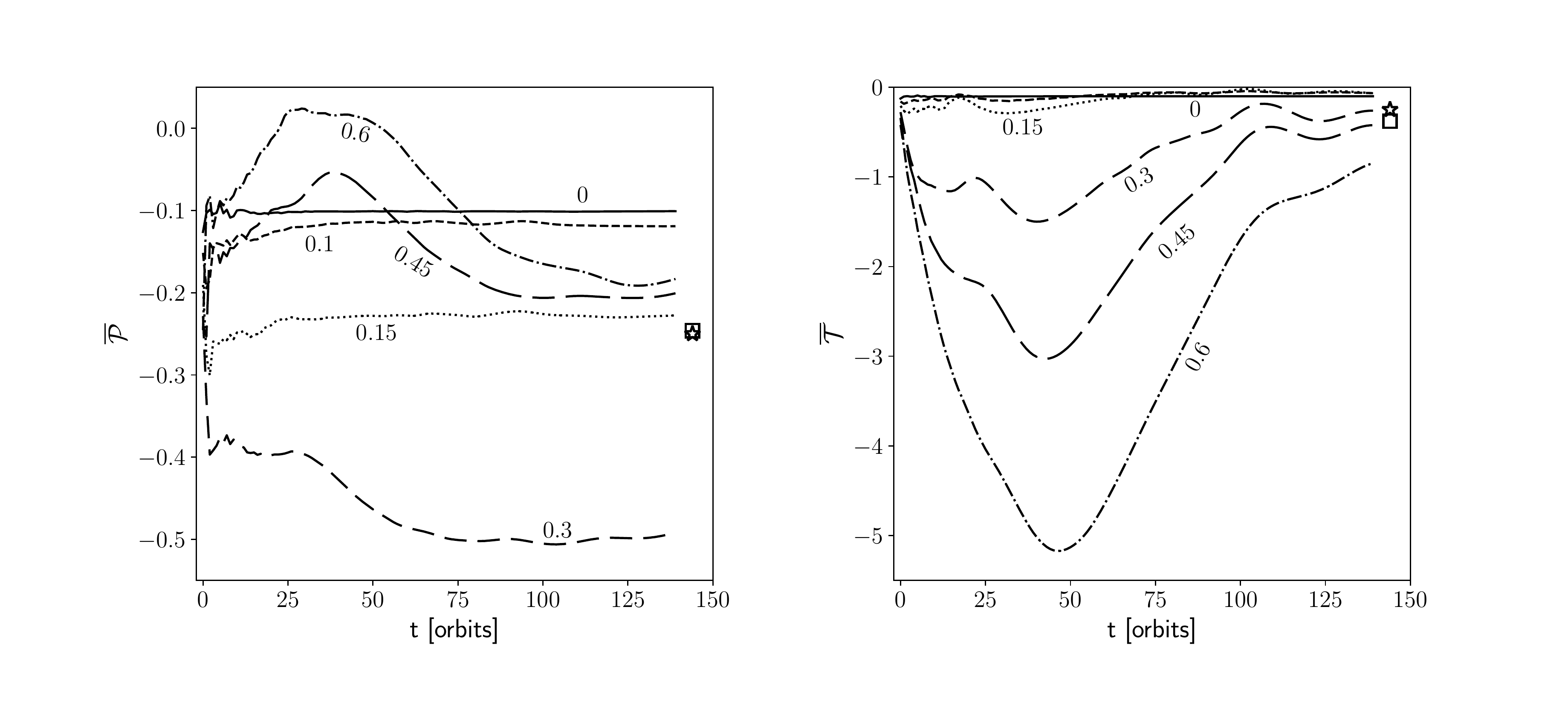}
  \caption{Same as Figure \ref{fig:diff_ecc} but for $\qmfive=50$. 
 }
\vskip 0.25cm
\label{fig:q2_diff_ecc}
\end{figure*}

\subsection{Models with $\alpha=0$ and $h=0.05$}
In this Section we assume that $h=0.05$ and $\alpha=0$, i.e. the unperturbed surface density is 
constant along $R$, so that $\Sigma_{t=0} = \Sigma_{0}=$const. From a numerical point of view, an 
initial constant surface density reduces spurious reflections in the boundaries and preserves reasonably 
well the mass in our computational box.

Retrograde perturbers excite tightly-wound density waves in the disk (see Figure \ref{fig:disk_map}).
The perturbers repeatedly catch their own wakes
with a frequency $2\omega$.  As a result, the surface density perturbation
$\Sigma-\Sigma_{0}$ is very complex, changing from positive to negative values in the
radial direction on a short spatial scale.

\begin{figure*}
\includegraphics[width=199mm,height=205mm]{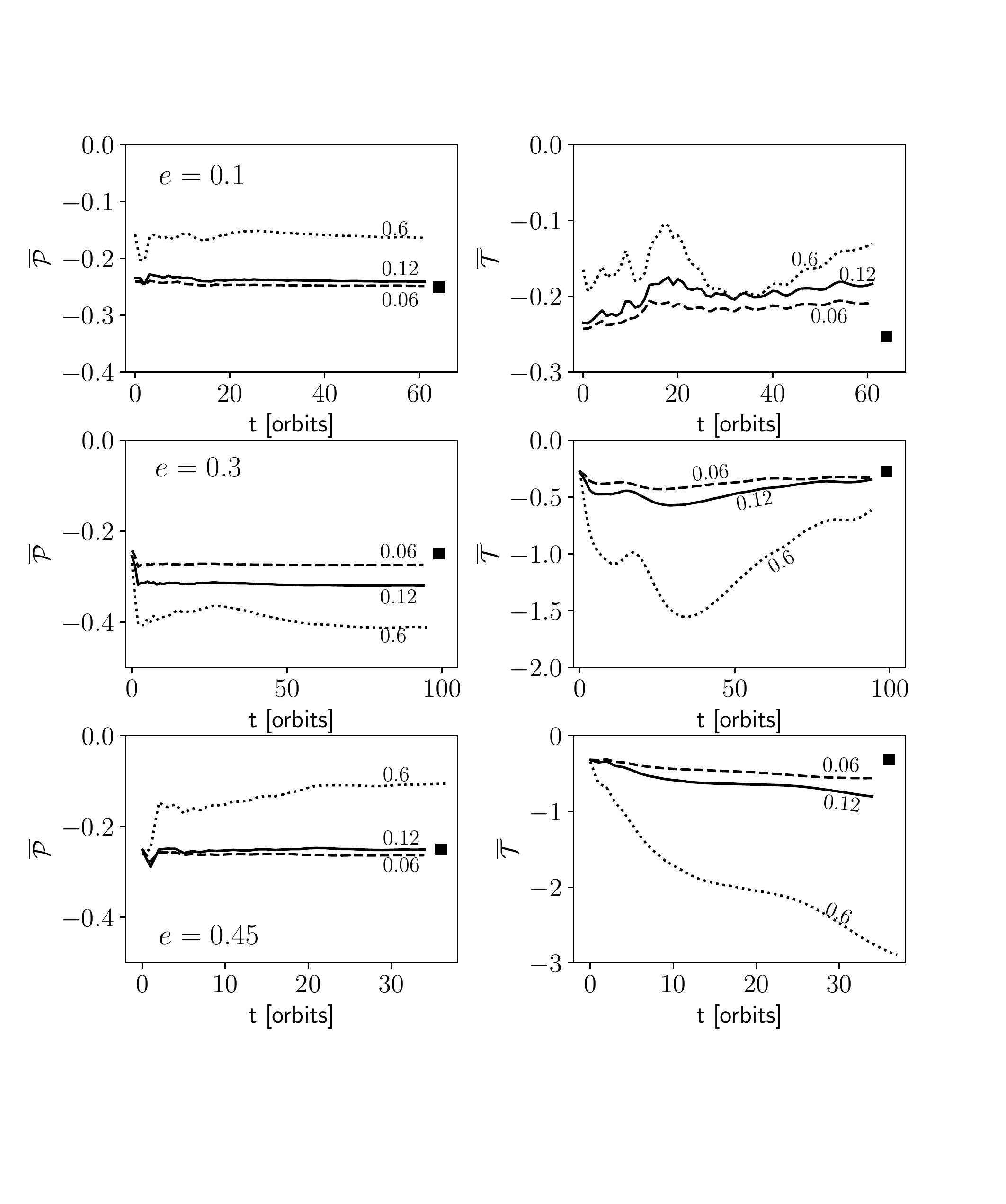}
  \caption{Evolution of the power (left panels) and torque (right panels)
for $e=0.1$ (top panels), $e=0.3$ (middle) and $e=0.45$ (bottom panels). Different curves are 
for different ${\mathcal{E}}$. The values of ${\mathcal{E}}$ are given at each curve.
The squares indicate the value predicted in the LA.
In all cases, the mass ratio is $\qmfive=1$.
 }
\vskip 0.1cm
\label{fig:diff_ecc_time}
\end{figure*}

\subsubsection{Models with ${\mathcal{E}}=0.6$}
\label{sec:soft_const}
In this Section we fix the values of $\alpha$, $h$ and ${\mathcal{E}}$ 
and study how the power and the torque depend on the orbital eccentricity. 
We take $\alpha=0$, $h=0.05$ and $\mathcal{E}=0.6$.
Figure \ref{fig:diff_ecc} shows $\overline{\mathcal{P}}$ and $\overline{\mathcal{T}}$ for a 
mass ratio $\qmfive=1$. We see
that $\overline{\mathcal{P}}$ remains fairly constant with time if $e\leq 0.3$. For $e\geq 0.45$, the shape of 
$\overline{\mathcal{P}}$ versus time is not so flat, having maxima and minima. 

Another remarkable feature is that
$\left<\overline{\mathcal{P}}\right>_{140}$, the mean value of $\overline{\mathcal{P}}$ 
between $t=5$ orbits and $t=140$ orbits, 
changes from $-0.1$ for $e=0$ to $-0.4$ for $e=0.3$ (see Figure \ref{fig:pw140}). 
For $e=0.6$, 
$\left<\overline{\mathcal{P}}\right>$ 
takes a similar value as for $e=0$. The LA predicts $\left<\overline{\mathcal{P}}\right>= -0.25$. 
Thus, for eccentricites around the end values of our interval, the measured values of the power are
a factor of $2.5$ smaller than the LA value. On the other hand, for eccentricities between
$0.15$ and $0.37$, the power in absolute value is larger than the LA value.
This is likely a consequence of the Lindblad resonant effects which are ignored in the LA. In fact, in the case of retrograde circular orbits, for which
there is no Lindblad resonances, the power is always less or equal to the LA value.

On the other hand, the curves $\overline{\mathcal{T}}$ versus time exhibit a deep valley at 
$t\simeq 30-50$ orbits for $e\geq 0.3$ (see right panel in Figure \ref{fig:diff_ecc}). 
In particular, in the case $e=0.6$, $|\overline{\mathcal {T}}|$ grows from $\sim 1.2$ at $t=5$ orbits 
to $\sim 5$ after $48$ orbits. These
values are much larger than the value predicted in the LA (which is $0.38$, see Figure \ref{fig:mean_pw_tq_LA}).
As long-term runs indicate (Figure \ref{fig:tq_long_runs}), the torque does not converge asymptotically
to a constant value, but shows large variations over the runtime of our simulations. 
Therefore, we cannot establish well-defined values of $\left<\overline{\mathcal{T}}\right>$, 
at least when ${\mathcal{E}}=0.6$. 

The temporal variations in $\overline{\mathcal{P}}$ but mainly in $\overline{\mathcal{T}}$ reflect the fact 
that the flow properties are not periodic functions of time (in this sense we say 
that the disk has not reached a ``steady state'').
If the evolution of the disk could be described through the
combination of linear density waves, it is expected that a steady state is reached in a few orbits. 
The temporal variations are a consequence of the secular evolution of the disk 
because of the deposition of angular momentum carried by the wake through shocks.
A steady-state will be reached on scales of the viscous time 
($t_{\nu}\simeq e^{2}a^{2}/\nu\simeq 5\times 10^{3}$ orbits, assuming $e=0.6$),
which is much longer than the crossing time. For perturbers in prograde and circular orbits, 
a description of the shock damping of waves in the weakly non-linear regime (low-mass perturbers)
can be found in \citet{goo01}. In this regime, inviscid linear theory still predicts
correctly the torques on the disk, although it implicitly assumes some dissipation.
Here we find that for extended perturbers with ${\mathcal{E}}=0.6$ in retrograde and eccentric
orbit, the magnitude of the torque is sensitive to 
the shock propagation and wave damping, even if the excitation of the wake is linear.

While the curves $\overline{\mathcal{P}}(t)$ and $\overline{\mathcal{T}}(t)$ should not depend on 
the adopted value of $q$ if the density waves induced in the disk were strictly linear, some dependence 
on $q$ can be expected in the presence of wave damping.
Figure \ref{fig:q2_diff_ecc} shows $\overline{\mathcal{P}}$ and $\overline{\mathcal{T}}$, 
as Figure \ref{fig:diff_ecc}, but for $\qmfive=50$.
The amplitude of the temporal variations of $\overline{\mathcal{P}}$ for $e=0.45$ and $e=0.6$ 
increases when $\qmfive$ is varied from $1$ to $50$.
$\overline{\mathcal{T}}(t)$ also changes in a comparable amount but they are less notorious because 
the fractional change is smaller.

\subsubsection{Varying the softening radius}
\label{sec:diff_soft}
One expects that the LA will become more accurate as ${\mathcal{E}}$ decreases, because the main
contribution to the drag force will arise from a closer vicinity of the body, admiting a local
description. This holds true for prograde eccentric orbits
\citep{san19}, as well as for retrograde circular orbits \citep{san18}.

\begin{figure}
\includegraphics[width=92mm,height=85mm]{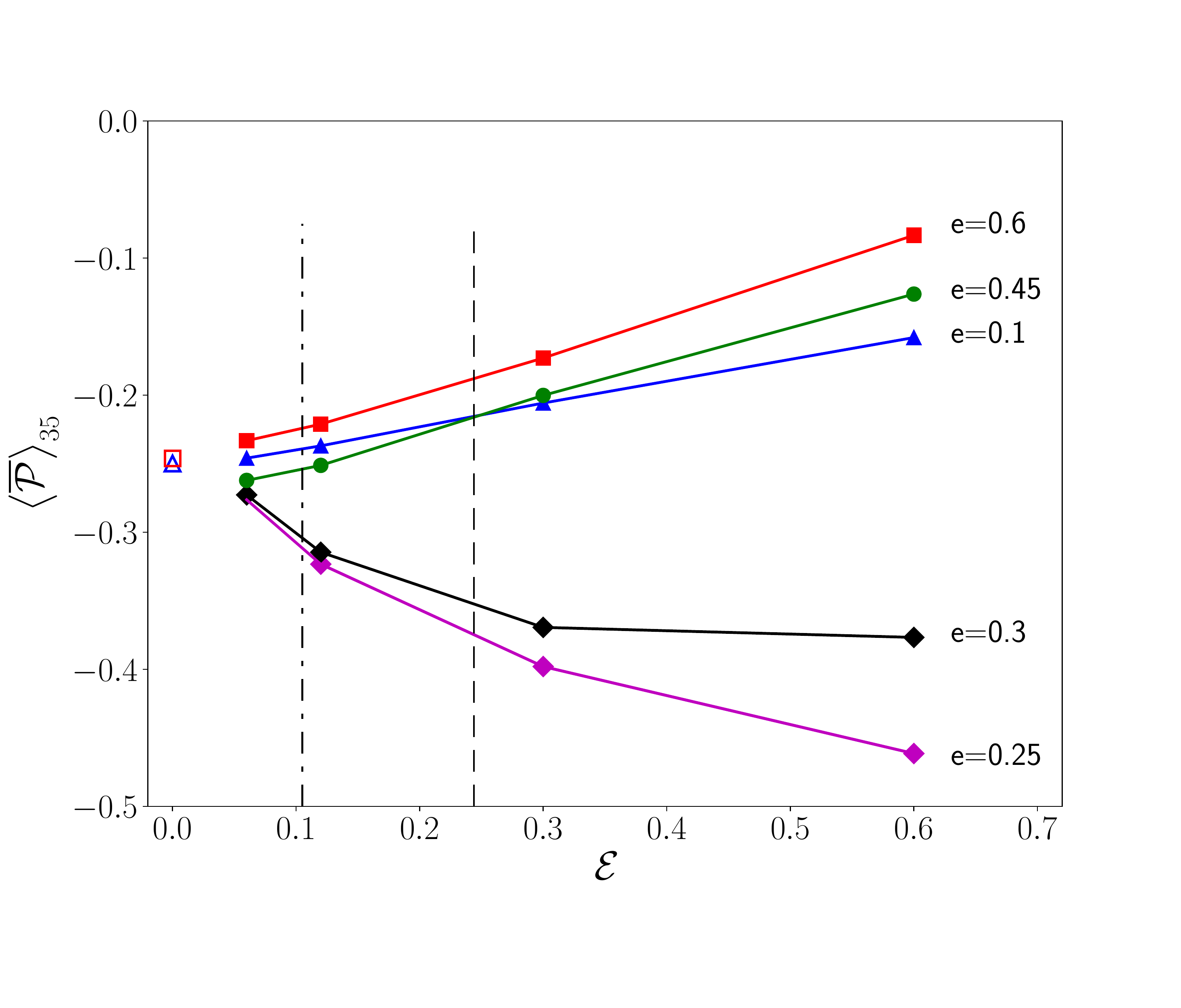}
  \caption{Mean power between the $5$th and $35$th orbits as a function of 
${\mathcal{E}}$ for 
simulations with different eccentricities. In these models $\alpha=0$, $h=0.05$ and $\qmfive=1$.
The vertical lines mark the value of $\mathcal{E}$ for which the error in the power
introduced by the LA is a factor $1.5$ (dashed line) or $1.25$ (dot-dashed line)
from the values measured in the simulations. The hollow symbols at the left side
of the Figure indicate the values
predicted in the LA.
 }
\vskip 0.25cm
\label{fig:compilation_power}
\end{figure}

\begin{figure}
\hskip -0.30cm
\includegraphics[width=99mm,height=85mm]{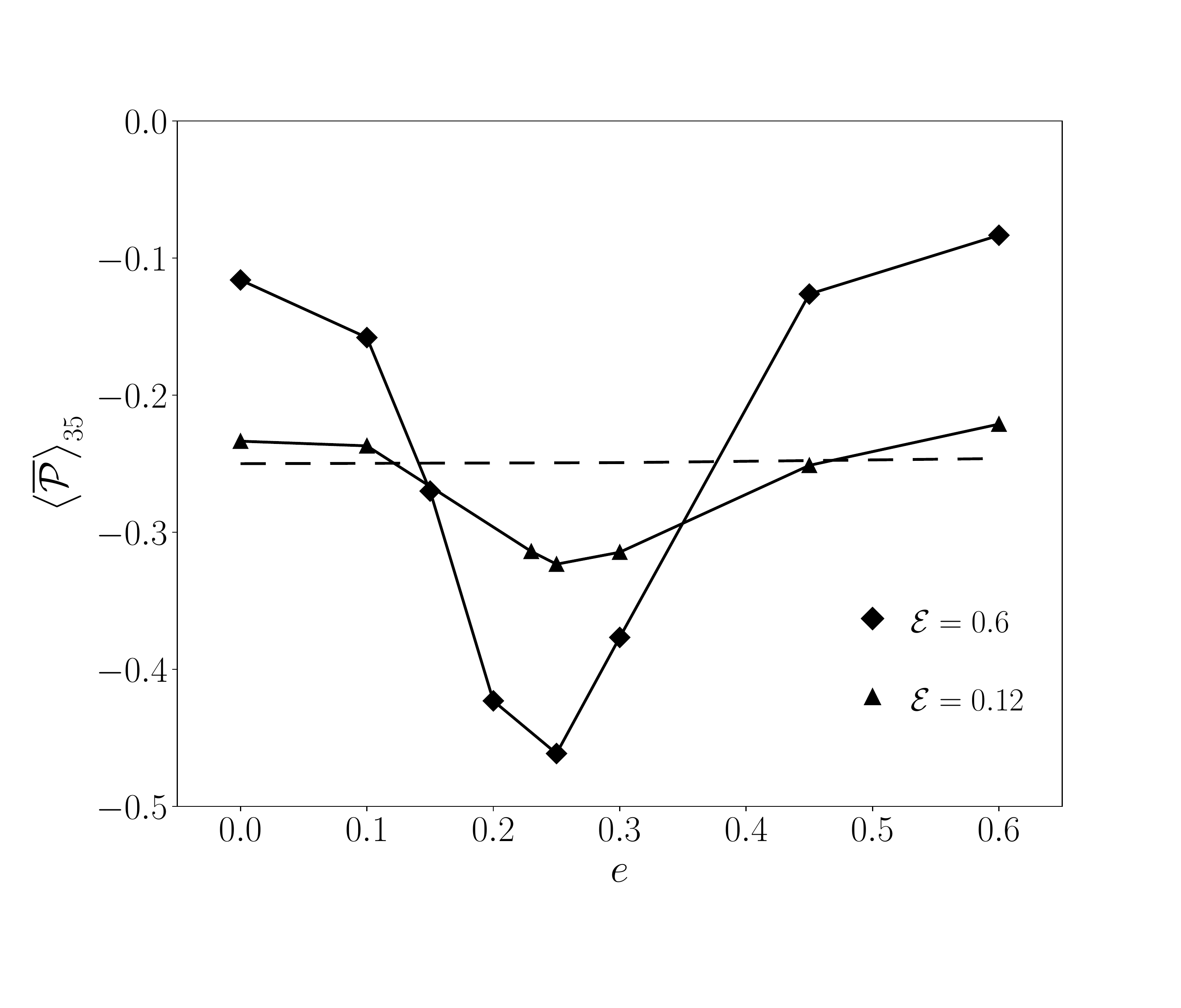}
\caption{$\left<\overline{\mathcal{P}}\right>_{35}$ versus eccentricity for ${\mathcal{E}}=0.6$ (diamonds) and for ${\mathcal{E}}=0.12$ (triangles). The predicted power
in the LA is also shown (dashed line). In these models $\alpha=0$, $h=0.05$ and $\qmfive=1$.}
\vskip 0.2cm
\label{fig:pw_h005_vs_ecc}
\end{figure}

Figure \ref{fig:diff_ecc_time} shows the power and the torque for ${\mathcal{E}}$ between $0.06$ and
$0.6$. For ${\mathcal{E}}\leq 0.12$, the power remains fairly constant over time. 
Indeed, the temporal behaviour of the power is already flat for ${\mathcal{E}}\simeq 0.3$ (not shown).
In addition, the value of the power converges (from below or from above) to the value predicted in the LA
as ${\mathcal{E}}$ decreases. This is more clearly seen in Figure \ref{fig:compilation_power}, 
where we plot $\left<\overline{\mathcal{P}}\right>_{35}$ as
a function of ${\mathcal{E}}$, where the brakets $\left<...\right>_{35}$ denote the 
mean value between $t=5$ and $t=35$ orbits. 
The choice of the values of $e$ in that Figure is not completely arbitrary.
We selected $e=0.25$ because, as already mentioned in \S \ref{sec:soft_const},
the maximum value of the power (in absolute value) occurs at this critical eccentricity.
This is more easily visualized in 
Figure \ref{fig:pw_h005_vs_ecc}, where we show $\left<\overline{\mathcal{P}}\right>_{35}$ versus eccentricity. The value $e=0.6$ was selected because the power reaches
its minimum value there (see Figure \ref{fig:pw_h005_vs_ecc}).

For our purposes, it is convenient to define ${\mathcal{E}}_{p;1.5}$ as the maximum value required 
for the LA to give the power within a factor of $\sim 1.5$ from the values measured in the simulations.
In other words, if ${\mathcal{E}}\leq \mathcal{E}_{p;1.5}$ then the ratio between the measured and
the predicted power lies between $0.66$ and $1.5$. We find that ${\mathcal{E}}_{p;1.5}=0.25$. 
For $\mathcal{E}=0.12$,  $\left<\overline{\mathcal{P}}\right>_{35}$ as measured in the simulations lies
between $-0.22$ and $-0.32$, in broad agreement with the value $-0.25$ derived in the LA.

\begin{figure}
\includegraphics[width=92mm,height=85mm]{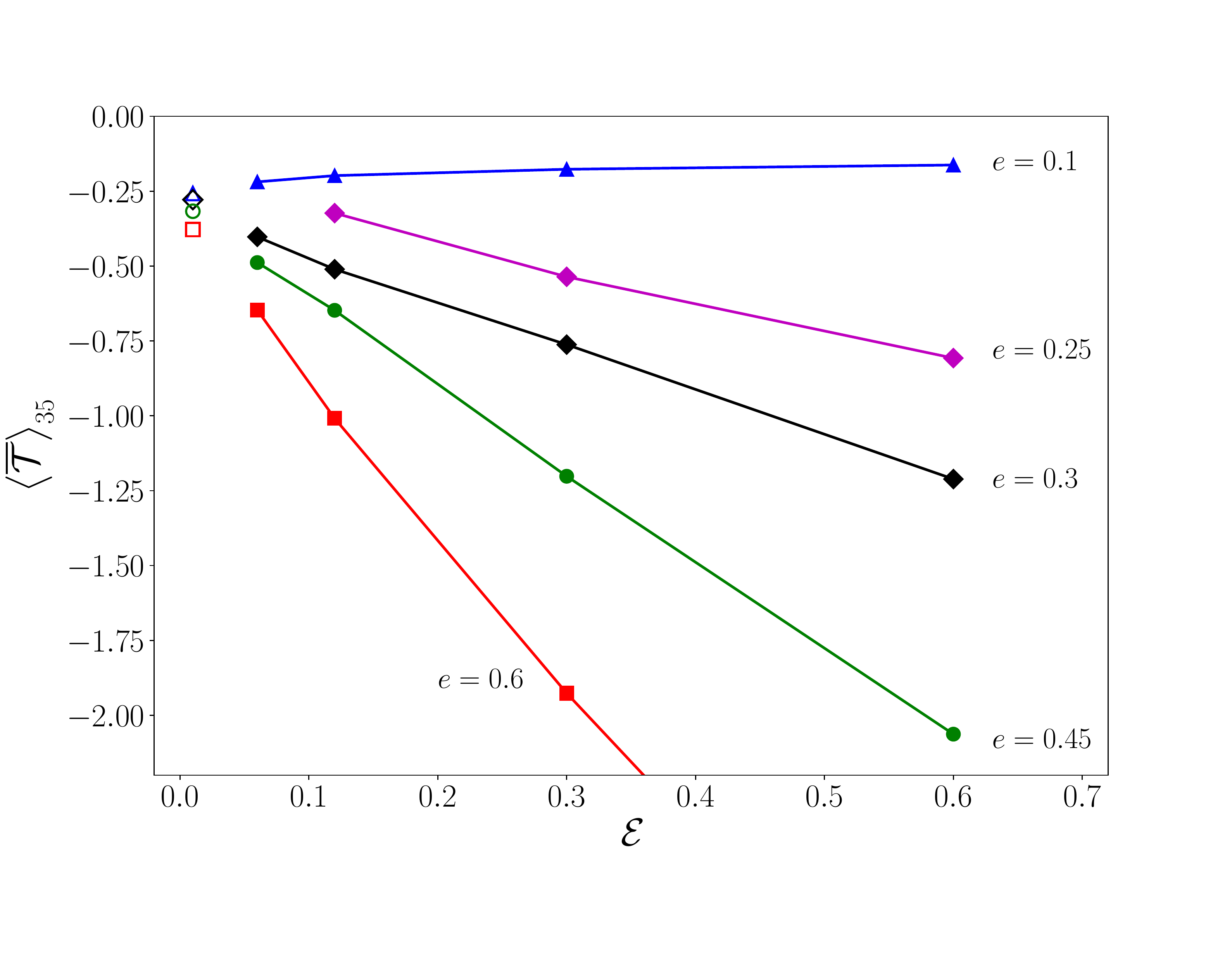}
  \caption{Same as Figure \ref{fig:compilation_power} but for the torque. For reference, the hollow symbols 
indicate the values expected using the LA. 
 }
\vskip 0.25cm
\label{fig:compilation_torque}
\end{figure}

Regarding the torque, the amplitude of its oscillations is reduced as ${\mathcal{E}}$ is taken smaller
(Figure \ref{fig:diff_ecc_time}).
For $e=0.45$, the amplitude of the temporal variations of the torque is still
comparable to its mean value even for ${\mathcal{E}}=0.06$.
For ${\mathcal{E}}=0.06$ and $e\leq 0.3$, the torque variations become relatively small. 
In these cases ($e\leq 0.3$ and ${\mathcal{E}}=0.06$), the discrepancy between the torque 
measured in the simulations and the predicted value in the LA is $\leq 25\%$
(see also Figure \ref{fig:compilation_torque}). Since $t_{e}$ depends
on the difference between $\overline{\mathcal{P}}$ and $\eta^{-1} \overline{\mathcal{T}}$ 
(see Eq. \ref{eq:te_B}), it remains uncertain to determine whether $e$ grows or damps in these cases.

Given that the torque may oscillate on a timescale $50-100$ orbits,  
$\left<\overline{\mathcal{T}}\right>_{35}$
should be interpreted with caution, as it reflects the depth of the first valley. Still, it is 
illustrative to see that $\left<\overline{\mathcal{T}}\right>_{35}$ converges to the value predicted in the 
LA as ${\mathcal{E}}$ decreases (Figure \ref{fig:compilation_torque}).
An extrapolation of the curves in Figure \ref{fig:compilation_torque} strongly suggests that
${\mathcal{E}}_{t;1.5}\simeq 0.04$, where ${\mathcal{E}}_{t;1.5}$ is the equivalent to ${\mathcal{E}}_{p;1.5}$
but for the torque.

\subsection{Varying $h$}
\label{sec:diff_h}

Our reference value for the aspect ratio, $h=0.05$, is representative for protoplanetary disks. 
For AGN accretion disks, the aspect ratio is less constrained, but models suggest a range for $h$ between
$0.01$ and $0.1$ \cite[e.g.,][]{sir03}.
Since the local Mach number for an object in retrograde orbit is $\simeq 2/h$,
the perturbed density in the disk depends on $h$. We have carried out a set of simulations 
with $h=0.025$ and $h=0.1$ (again with $\alpha=0$), to check how 
the results depend on $h$.

\begin{figure}
\hskip -0.30cm
\includegraphics[width=99mm,height=85mm]{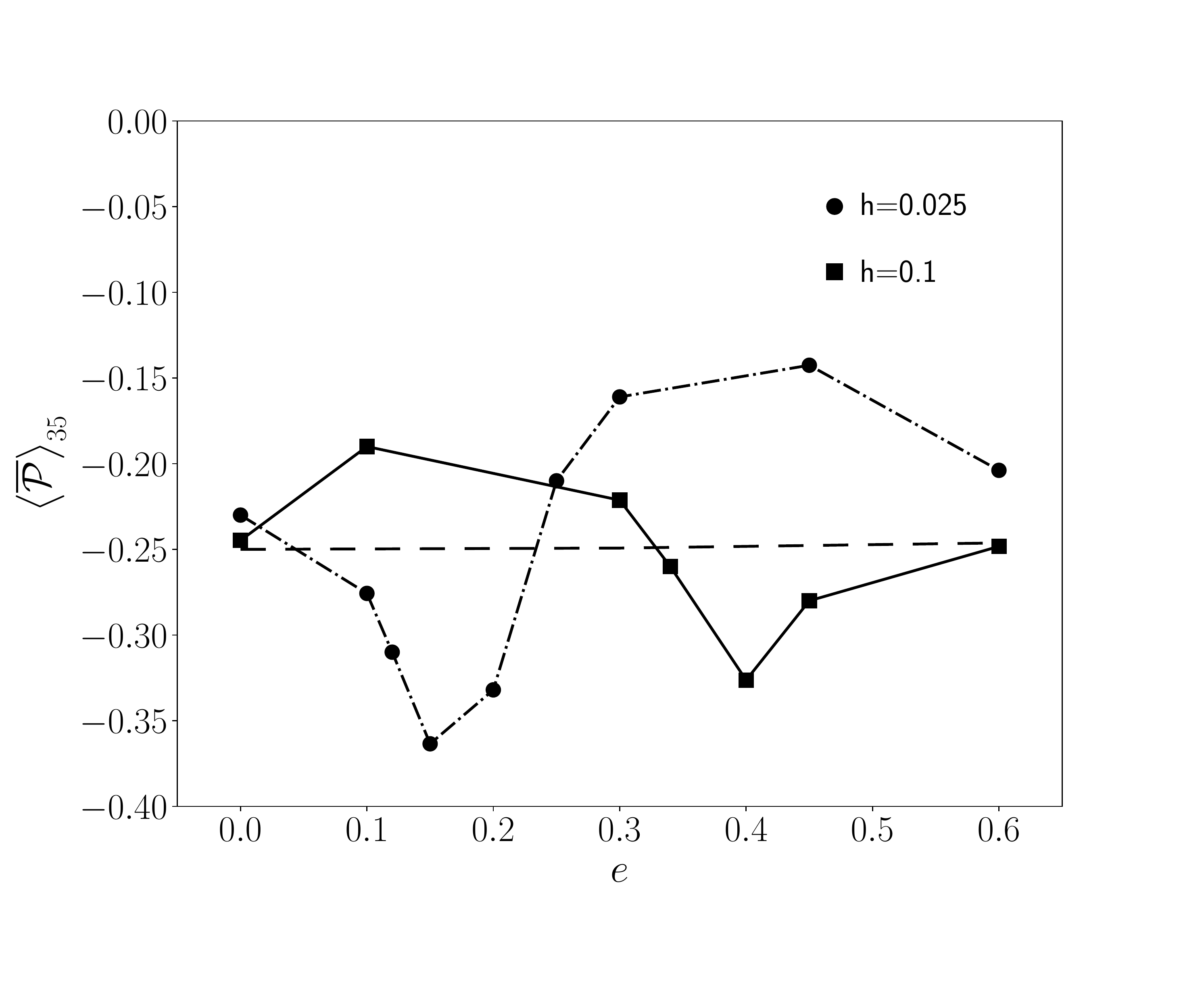}
\caption{$\left<\overline{\mathcal{P}}\right>_{35}$ as a function of eccentricity 
for $h=0.025$ (circles) and $h=0.1$ (squares). In both cases, $\alpha=0$ and 
${\mathcal{E}}=0.12$. The power in the LA is indicated by the dashed line.}
\vskip 0.2cm
\label{fig:ecc_crit}
\end{figure}

Figure \ref{fig:ecc_crit} plots the mean power $\left<\overline{\mathcal{P}}\right>_{35}$
as a function of eccentricity. We see that
the critical eccentricity depends on $h$. The critical eccentricity is $0.4$ for $h=0.1$
and $0.15$ for $h=0.025$.

\begin{figure*}
\includegraphics[width=199mm,height=89mm]{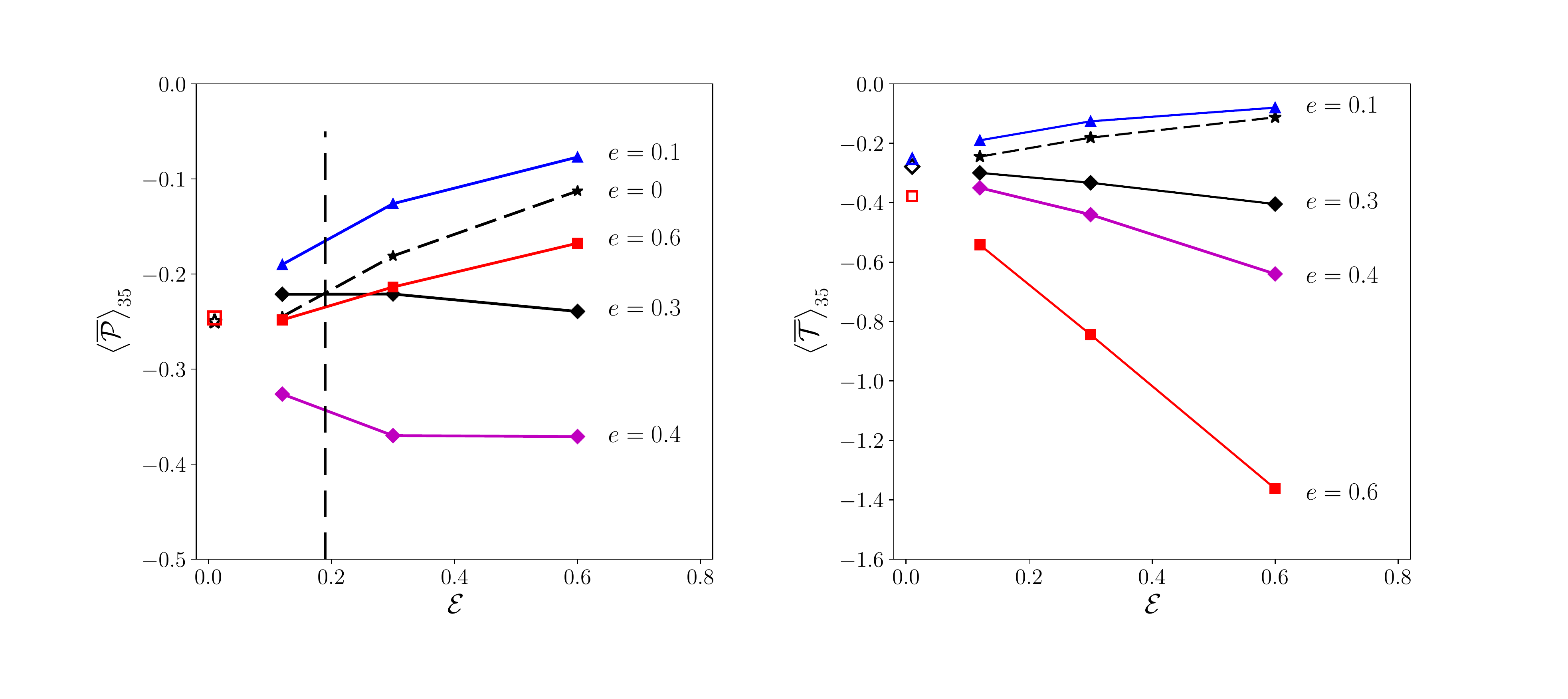}
  \caption{$\left<\overline{\mathcal{P}}\right>_{35}$ (left panel) and  
$\left<\overline{\mathcal{T}}\right>_{35}$ (right panel), as a
function of ${\mathcal{E}}$, for $h=0.1$ and different eccentricities. The hollow symbols indicate the value
predicted in the LA. The dashed line marks $\mathcal{E}_{p;1.5}$. In all cases, we take $\qmfive=1$.
 }
\vskip 0.25cm
\label{fig:compilation_h01}
\end{figure*}

\begin{figure}
\hskip -0.3cm
\includegraphics[width=92mm,height=82mm]{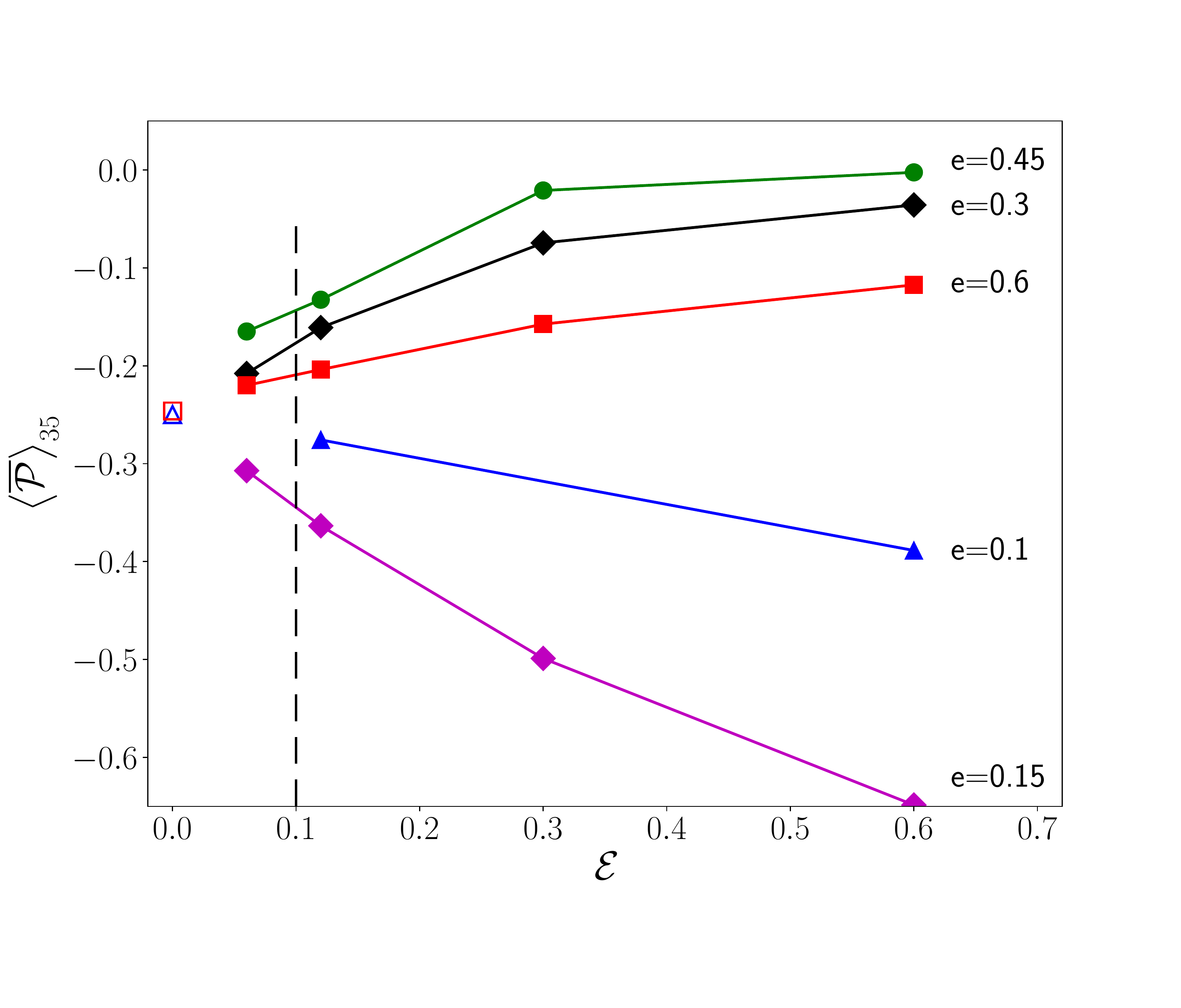}
  \caption{Mean power, $\left<\overline{\mathcal{P}}\right>_{35}$, as a
function of ${\mathcal{E}}$, for $h=0.025$ and different eccentricities. The dashed line indicates
$\mathcal{E}_{p;1.5}$. The empty symbols at the left side mark the predicted
values in the LA. In all cases, we take $\qmfive=1$.
 }
\vskip 0.05cm
\label{fig:compilation_power_h0025}
\end{figure}

Figure \ref{fig:compilation_h01} shows
$\left<\overline{\mathcal{P}}\right>_{35}$ and $\left<\overline{\mathcal{T}}\right>_{35}$
for $h=0.1$.  
$\left<\overline{\mathcal{P}}\right>_{35}$ 
presents a dispersion around the LA values similar to that found for $h=0.05$.
We find that ${\mathcal{E}}_{p;1.5}=0.2$, which is similar to the value found for $h=0.05$.

For $h=0.1$, the values of  $\left<\overline{\mathcal{T}}\right>_{35}$ get closer to the LA estimates  
than for $h=0.05$. We infer ${\mathcal{E}}_{t;1.5}\simeq 0.05$.
For $e\leq 0.4$ and ${\mathcal{E}}=0.12$, we run the simulations until $120$ orbits and found that 
$\left<\overline{\mathcal{P}}\right>$ and $\left<\overline{\mathcal{T}}\right>$ are similar, implying
that $\mathcal{B}$ is significantly smaller than $\mathcal{P}$.

Figure \ref{fig:compilation_power_h0025} shows the power for $h=0.025$. Interestingly,
the values of $\left<\overline{\mathcal{P}}\right>_{35}$ spread apart from the values 
derived in the LA. In particular, we notice that the power is minimum (in absolute value)
at $e=0.45$ (i.e. $e/e_{\rm crit}=3$). If we restrict ourselves to orbital eccentricities $0\leq e\leq 2e_{\rm
crit}=0.3$, we obtain ${\mathcal{E}}_{p;1.5}=0.12$.

\begin{figure}
\hskip -0.3cm
\includegraphics[width=92mm,height=80mm]{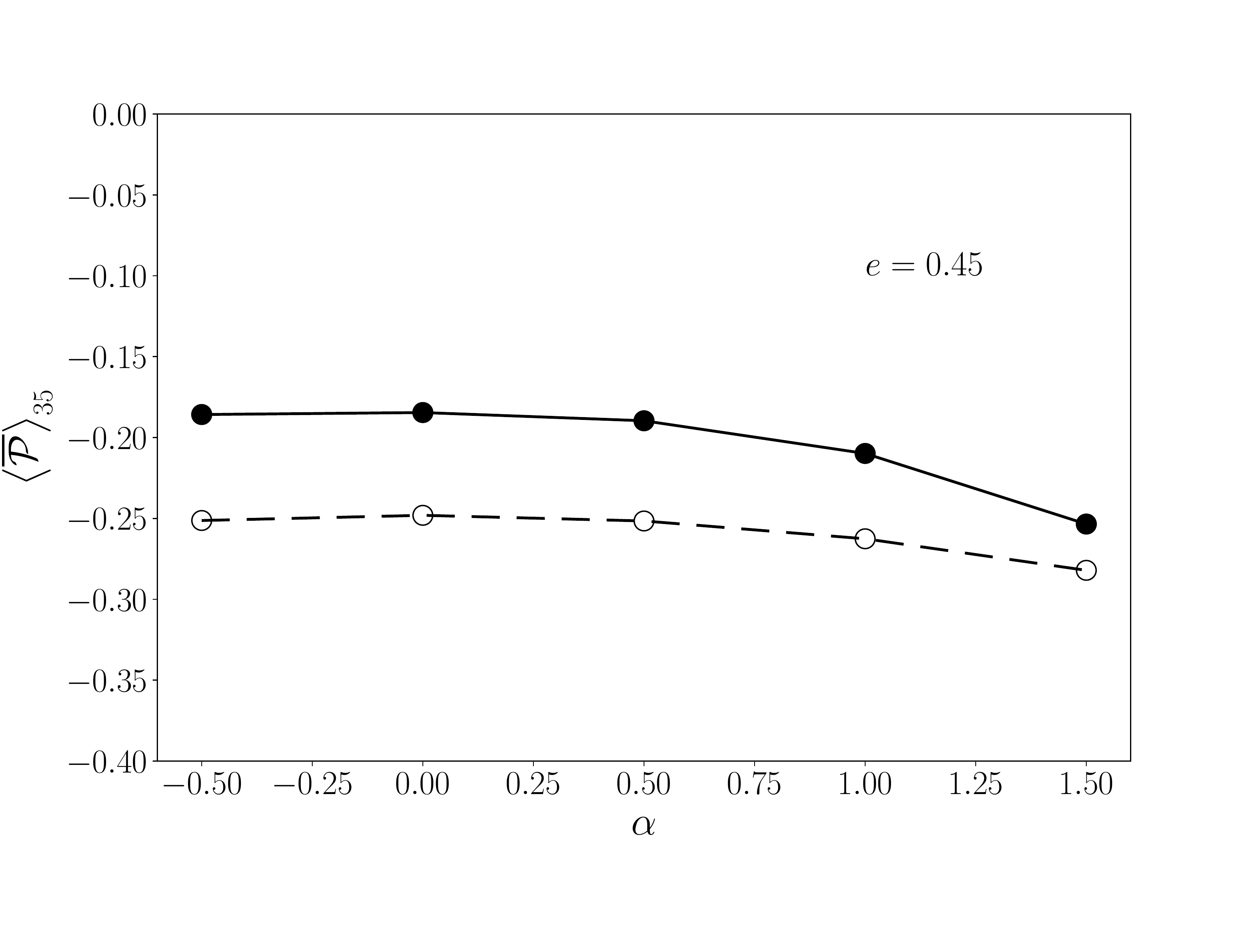}
  \caption{Mean values of the power, $\left<\overline{{\mathcal{P}}}\right>_{35}$, as a
function of $\alpha$ for $e=0.45$ and ${\mathcal{E}}=0.3$ (solid line). 
The dashed line with hollow circles marks the value
predicted in the LA. In all cases, we take $\qmfive=10$.
 }
\vskip 0.25cm
\label{fig:compilation_diff_alpha}
\end{figure}

Putting together the results obtained for $h=0.025, 0.05$ and $0.1$, we find the following
rules of thumb. 
The absolute value of the power is maximum at a critical eccentricity
\begin{equation}
e_{\rm crit}\simeq 0.25 \left(\frac{h}{0.05}\right)^{2/3}.
\end{equation} 
At eccentricities $\simeq e_{\rm crit}$, the power is larger, in absolute value, than what the LA predicts.
The LA predicts the power, with an error less than $30\%$,
at eccentricities around $e\simeq 0.6e_{\rm crit}$ provided that $\mathcal{E}\leq 0.6$.
On the other hand, if we take ${\mathcal{E}}\leq 0.12$, the LA predicts the power in
the range $e\leq 2e_{\rm crit}$ within a factor less than $1.5$.
Finally, our results suggest that $\mathcal{E}_{p;1.5}= 0.2\,{\rm min}[1, (h/0.05)]$
in the range of eccentricities $0<e<0.6$.

\subsection{Varying $\alpha$}
\label{sec:diff_alpha}
We have run models with $\qmfive=10$, $h=0.05$, $e=0.45$ and ${\mathcal{E}}=0.3$, and
different $\alpha$.
Figure \ref{fig:compilation_diff_alpha} shows that the local approximation is equally well 
regardless the value of $\alpha$.
This is expected because the unperturbed surface density changes on a radial scale of 
$|\Sigma_{0}/d\Sigma_{0}/dR|\simeq a/|\alpha|$, which is much larger than the length of the wake that 
contributes most to the drag force. 

\section{Implications for the evolution of compact objects in 3D accretion disks}
\label{sec:compactcase}
Our simulations consider the response of a 2D disk to a softening potential, ignoring mass accretion
onto the perturber.
In real life, COs, such as BHs or neutron stars,
have very small or null physical radii and they are embedded in disks
with finite scaleheight. Following \citet{san19}, we extend the results to the latter scenario.

Suppose that the LA predicts the power or the torque with some permissible error if
$R_{\rm soft}\leq \tilde{R}_{\rm soft}\equiv {\mathcal{E}}_{\rm max}H$. Once we know $\tilde{R}_{\rm soft}$ in a 2D disk, 
denoted by $\tilde{R}_{\rm soft}^{(2D)}$, we can obtain $\tilde{R}_{\rm soft}^{(3D)}$, the maximum softening 
radius in a 3D disk.
In fact,  Figure 13 in \citet{san19} shows the relationship between $\mathcal{E}_{\rm max}^{(2D)}$
and $\mathcal{E}_{\rm max}^{(3D)}$.
In particular, for $h\geq 0.025$ and $e\leq 0.6$, we have found in
Section \ref{sec:diff_soft} that ${\mathcal{E}}^{(2D)}_{p;1.5}=0.1$.
This translates into $\mathcal{E}^{(3D)}_{p;1.5}=0.01$ for extended perturbers embedded in 3D disks. 
Note that $\mathcal{E}_{p;1.5}^{(3D)}$ is smaller than $\mathcal{E}_{p;1.5}^{(2D)}$ because the
drag force depends logarithmically on $R_{\rm soft}^{-1}$ in a 3D disk, while it scales as
$R_{\rm soft}^{-1}$ in a 2D disk.

For point-like objects, like COs, the local approximation will be valid as long as the accretion radius $R_{\rm acc}$ 
(which is the minimum effective scale of the interaction) is smaller than $\tilde{R}_{\rm soft}^{(3D)}$.
Note that $R_{\rm acc}$ of a body in eccentric orbit may vary along the orbit, being
maximum at apocenter because the relative velocity is minimum.  Thus, if we 
demand $R_{\rm acc}\leq \tilde{R}_{\rm soft}$ at apocenter, we can derive
an upper limit on the mass ratio of the inspiral. Using $R_{\rm acc}=2GM_{p}/V_{\rm rel}^{2}$,
with $V_{\rm rel}\simeq 2\omega a/\sqrt{1+e}$ at apocenter, the above condition can be cast, in
terms of $q$, as $q\leq 2h \mathcal{E}_{\rm max}$ for COs.

For a disk with $h\geq 0.025$ and for $\mathcal{E}_{p;1.5}^{(3D)}=0.01$ (see above), we obtain
that the LA predicts the power and therefore also the rate of inspiral $t_{a}$, within a factor of $1.5$,
for inspirals having $\qmfive\leq 50$.
Interestingly, this range of masses includes EMRIs (see \S \ref{sec:basic}).
We highlight that the estimates of the power and $t_{a}$ are robust
in the sense that they are weakly dependent on the orbital eccentricity.

\begin{figure*}
\includegraphics[width=199mm,height=85mm]{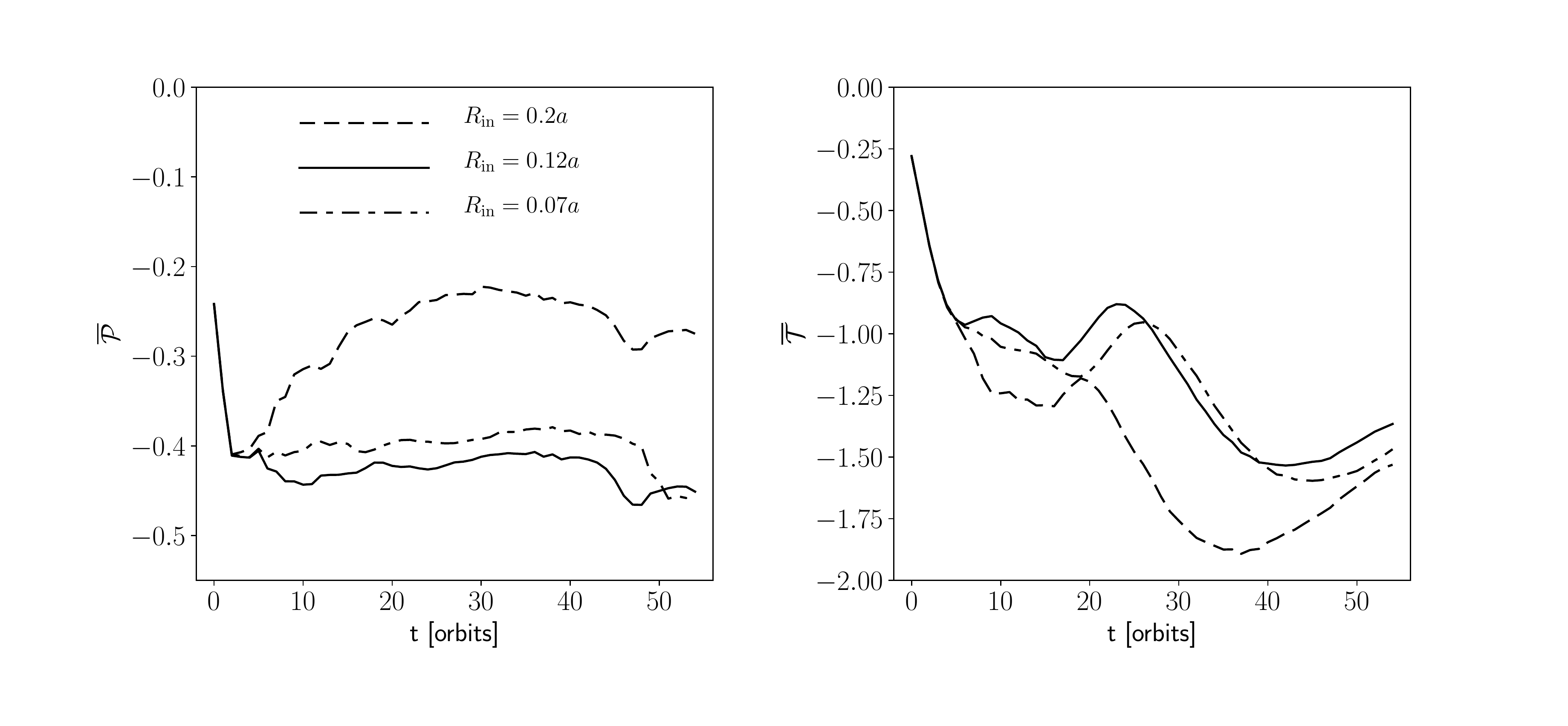}
  \caption{Evolution of the power (left panel) and torque (right panel) 
for different $R_{\rm in}$ in simulations with reflecting boundaries
(without waves-damping
zones). We took $\qmfive=5$, $e=0.3$, ${\mathcal{E}}=0.6$
and $R_{\rm out}=5.2a$ in all cases.
 }
\vskip 0.25cm
\label{fig:diff_rin_WALL}
\end{figure*}

The condition for the LA to predict the torque with the same error is much more restrictive.
For instance, consider a disk with $h=0.05$. For $e\leq 0.6$, our simulations suggest that 
$\mathcal{E}^{(2D)}_{t;1.5}\simeq 0.04$ (\S \ref{sec:diff_soft}). This value corresponds to
$\mathcal{E}^{(3D)}_{t;1.5}\simeq 10^{-4}$, implying $\qmfive\lesssim 1$. If we are only interested
in orbital eccentricities smaller than $0.3$, the corresponding condition is $\qmfive\lesssim 8$.

Owing that $t_{e}$ is inversely proportional to $|\overline{B}|=|\overline{P}-\eta^{-1}\overline{T}|$ (see Eq. 
\ref{eq:taue}), 
$t_{e}$ remains very unconstrained unless $\overline{P}$ and $\overline{T}$ are very dissimilar. 
In numerical simulations, it is difficult to determine the net evolution of the eccentricity 
because of alternating periods during which the eccentricity grows or damps.

\section{Conclusions}
\label{sec:conclusions}
A DF approach is commonly used to model the gravitational interaction between an accretion
disk and an orbiter moving on an eccentric or/and inclined orbit. 
This approach assumes that the interaction is local, i.e. the
gas ahead of the perturber remains unperturbed and, in addition, most of the contribution to the 
tidal forces arises from a region so close to the perturber that curvature terms are unimportant.
In this paper, we have considered the orbital evolution 
of a low-mass perturber, having an eccentricity between $0$ and $0.6$, and
an inclination of $180^{\circ}$, i.e. coplanar but retrograde orbit with respect to the gas disk.
In such a situation, the perturber moves supersonically with Mach numbers $40-200$ relative to the 
local gas, and it excites spiral waves that are wound tightly.

Notably, in typical accretion disk models, the local DF approach predicts that
the eccentricity is excited.
Nevertheless, in disks with a surface density that decays in the radial direction,
the timescale for the growth of the eccentricity is larger than the timescale for
the radially inward migration. Consequently,
there exists the possibility that the inspiral could merge with a non-zero eccentricity.
The DF approach also predicts that the rate of inspiral hardly depends on the orbital eccentricity.
The purpose of this work was to assess when the local DF approximation can be applied to
retrograde EMRIs.

We have computed the torque and the rate of work on a perturber on a fixed eccentric orbit
in 2D simulations.
The rate of energy loss by the perturber determines the rate of
inspiral, whereas a combination of the power and the torque determines the evolution of the
eccentricity.

We find that for eccentricities around the critical value
$e_{\rm crit}\simeq 0.25 (h/0.05)^{2/3}$, the power (in absolute value) is larger than predicted
in the LA. 
Nevertheless, the orbital-averaged power and torque converge to the values predicted by
the LA when ${\mathcal{E}}$ tends to zero. This reflects the fact that curvature effects and
resonances are less important for smaller bodies. 
For $h$ between $0.025$ and $0.1$, and for ${\mathcal{E}}\leq 0.1$, 
the LA predicts the power measured in 2D simulations within a factor of $1.5$ or less.
This condition for ${\mathcal{E}}$, which was found for extended perturbers embedded in 2D disks,
translates into $\qmfive \leq 50$ for COs in 3D disks. This mass range includes EMRIs.   

Numerical determinations of the mean torque require long-term simulations because the 
torque exhibits temporal variations, unless ${\mathcal{E}}$ is taken very small.  
Such long-term simulations are a numerical challenge because of the spurious noise
introduced through the boundaries. An extrapolation of our results indicates that
the LA estimates of the torque are within a factor $1.5$ of the measured values if
softened perturbers embedded in 2D disks with $h\geq 0.05$ have $\mathcal{E}<0.04$.
This implies $\qmfive\leq 1$ for COs embedded in 3D disks. However, we should stress
that even if the power and the torque are determined within a factor of $1.5$, the
error in the estimate of $t_{e}$ using the LA might be larger because it depends on
the difference $\overline{P}-\eta \overline{T}$.

\begin{figure*}
\includegraphics[width=199mm,height=85mm]{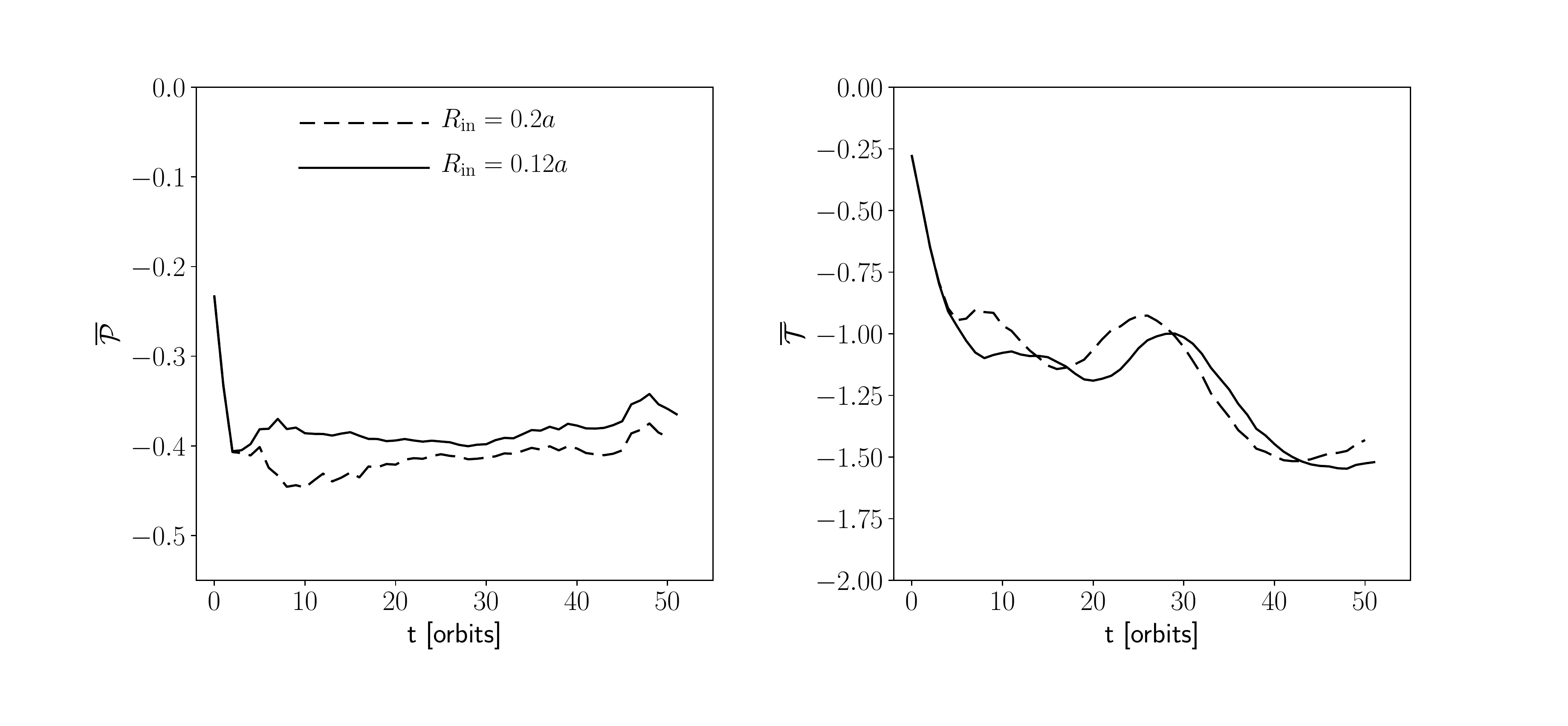}
  \caption{Same as Figure \ref{fig:diff_rin_WALL} but for open boundaries
(without wave damping zones). We took $\qmfive=10$, $e=0.3$, ${\mathcal{E}}=0.6$
and $R_{\rm out}=5.2a$.
 }
\vskip 0.25cm
\label{fig:diff_rin_OPEN}
\end{figure*}

\acknowledgments
I am grateful to Ra\'ul O. Chametla and Fr\'ed\'eric Masset for useful discussions,
and the anonymous referee for insightful comments.
The author acknowledges financial support by PAPIIT project IN111118.
The simulations were performed using the computers Tycho (Posgrado de Astrof\'{\i}sica-UNAM,
Instituto de Astronom\'{\i}a-UNAM and PNPC-CONACyT).

\appendix
\section{A. Boundary conditions and radial extent of the computational domain}
\label{sec:BC}
The finite size of the computational domain may induce undesirable phenomena,
leading to an inaccurate result. 
The domain should be taken as large as possible and the boundary
conditions should be chosen with the aim of minimizing spurious effects. 
Trusted results should be robust to reasonable changes of the
size of the box domain. In this Appendix, we study the sensitivity of the results
to the location of the inner edge of the computational domain and on the adopted inner 
boundary condition.  The outer edge is less problematic because it can
always be placed so far away that its effects are comparably less important.
We mainly focus on cases with the largest value of $\mathcal{E}$
(i.e. $\mathcal{E}=0.6$) because the effects of the inner boundary are more prominent
as $\mathcal{E}$ increases. In addition, we set up $\alpha=0$ and $h=0.05$ in all the 
simulations presented in this Appendix.

\begin{figure*}
\includegraphics[width=199mm,height=85mm]{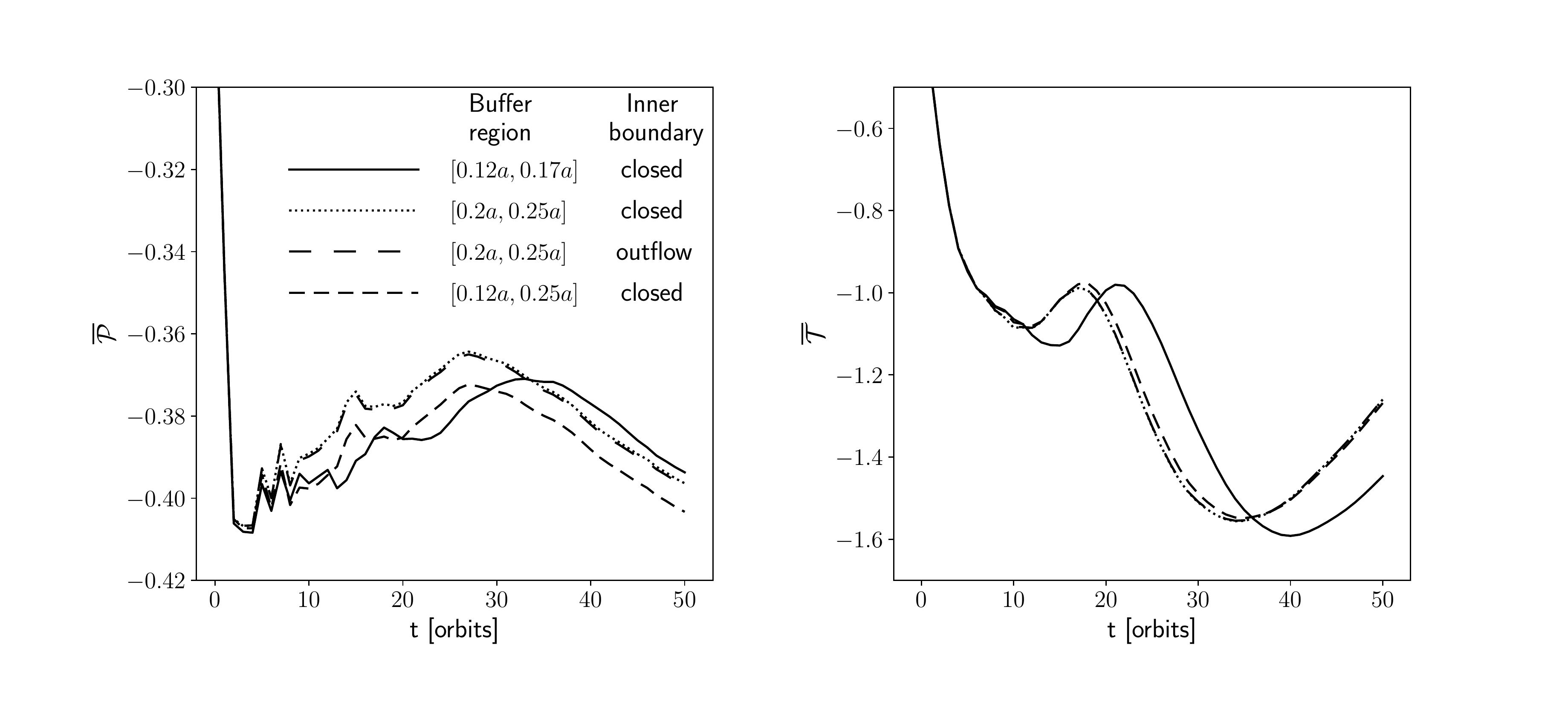}
  \caption{Evolution of the power (left panel) and torque (right panel) 
for different extents of the buffer region in the inner boundary. 
We used $\qmfive=1$, $e=0.3$, ${\mathcal{E}}=0.6$
and $R_{\rm out}=5.2a$,
in all cases. The buffer ring in the outer boundary is $R\in [4.95a, 5.2a]$.
 }
\vskip 0.25cm
\label{fig:diff_rin_stockh}
\end{figure*}

In the lack of resonances or collective modes, spurious
phenomena generated at the boundaries are usually diminished when the extent
of the domain increases. To illustrate this, Figure \ref{fig:diff_rin_WALL} shows 
$\overline{{\mathcal{P}}}$ and 
$\overline{{\mathcal{T}}}$ in simulations in which the boundaries
behave as a rigid wall, where large reflections are expected.
We take $R_{\rm out}=5.2a$, and three different values of $R_{\rm in}$ 
($0.07a, 0.12a$ and $0.2a$). 
In the two runs having $R_{\rm in}\leq 0.12a$, the power is approximately 
constant over time between $3t_{\rm orb}$ and $40t_{\rm orb}$.
The mean value of the power between the $5$th and $35$th orbits, 
$\left<\overline{{\mathcal{P}}}\right>_{35}$, in these two runs agrees
within $7\%$.

In these simulations, the power is not perfectly smooth but presents some wiggles, 
with a small amplitude of $\sim 4\%$. These small oscillations
are the consequence of the combination of two effects: (1) reflections at the inner boundary and 
(2) the interaction of the perturber with its own wake ahead of it, which has memory that the 
perturber was turned on suddenly at $t=0$.
The timescale of the fluctuations caused by the memory effect is very small 
(the timescale is $t_{\rm orb}$),
and the amplitude of these wiggles is attenuated, especially at early times, if the perturber is 
inserted slowly in the simulations (see Appendix \ref{sec:adiab_perturber}). 
Reflections at the inner boundary, on the other hand, lead to temporal variations with a frequency determined by the sound-crossing times $t_{s}$, defined as the time in which a sound wave 
takes to travel from $R=a$ to the inner boundary, and getting back after reflection.
In the run with $R_{\rm in}=0.07a$ we have that
$t_{s}\simeq 4.2t_{\rm orb}$, whereas 
$t_{s}\simeq 3.9t_{\rm orb}$ in the run with $R_{\rm in}=0.2a$. 
This small difference
in $t_{s}$ indicates that the wiggles caused by reflections are quite difficult to suppress
just by reducing $R_{\rm in}$.

On the other hand, $\overline{{\mathcal{T}}}$ presents a local maximum
 and then a local minimum
 (see right panel in Figure \ref{fig:diff_rin_WALL}). The minimum occurs about
$\sim 15t_{\rm orb}$ after the maximum. 
The locations of the maximum and minimum are not the same in the three simulations.
The torque measured in the simulation with $R_{\rm in}=0.07a$ is similar in shape to
the torque in the simulation using $R_{\rm in}=0.12a$, though slightly shifted horizontally.
The shift of about $3 t_{\rm orb}$ is much larger than the difference in $t_{s}$ between the
two simulations, which is only $\sim 0.2 t_{\rm orb}$. In fact, the local maxima and minima in
the torque are the result of global modes in the disk 
and cannot be intepreted in terms of boundary reflections which produce
variations on a shorter timescale. The temporal shift in the torque does not affect
much its mean value: the difference in the value of
$\left<\overline{{\mathcal{T}}}\right>_{35}$ between these
two simulations is less than $4\%$.

In addition to reflections, closed boundary conditions have the shortcoming that
gas is piled up in the inner boundary because it cannot leave
the computational domain, which is not realistic.
Alternatively open boundary conditions, which allow inflow and outflow through
the boundaries, can be considered.
The results of applying this condition at the inner and outer boundaries are shown in 
Figure \ref{fig:diff_rin_OPEN}. A weakness of using open boundaries is that the mass
in the disk is not preserved.

To overcome the limitations of closed and open boundary conditions, it is common
to implement buffer zones, where the density and velocity components are forced to gradually
back to their unperturbed values, as described in de Val-Borro et al. (2006). The timescale
of this damping process is proportional to the local dynamical timescale. The resultant power
and torque using wave killing regions with different widths are depicted in 
Figure \ref{fig:diff_rin_stockh}. The wave-damping ring extends from $R_{\rm in}$ up
to $R_{\rm end}$. The three simulations with $R_{\rm end}=0.25a$ yield similar results,
regardless the adopted boundary condition (closed or outflow).
In fact, the buffer regions are
large enough as to damp the waves before they reach $R=R_{\rm in}$, so that the results
only depend on $R_{\rm end}$; they do not depend either on $R_{\rm in}$ or on the
boundary condition. It is apparent that the torque for $R_{\rm end}=0.17a$ is shifted
to the right by $5t_{\rm orb}$, but this shift has a minor effect on mean torque:
$\left<\overline{{\mathcal{T}}}\right>_{35}$ agree within $6\%$ in the four
simulations shown in Figure \ref{fig:diff_rin_stockh}.

\begin{figure*}
\includegraphics[width=199mm,height=145mm]{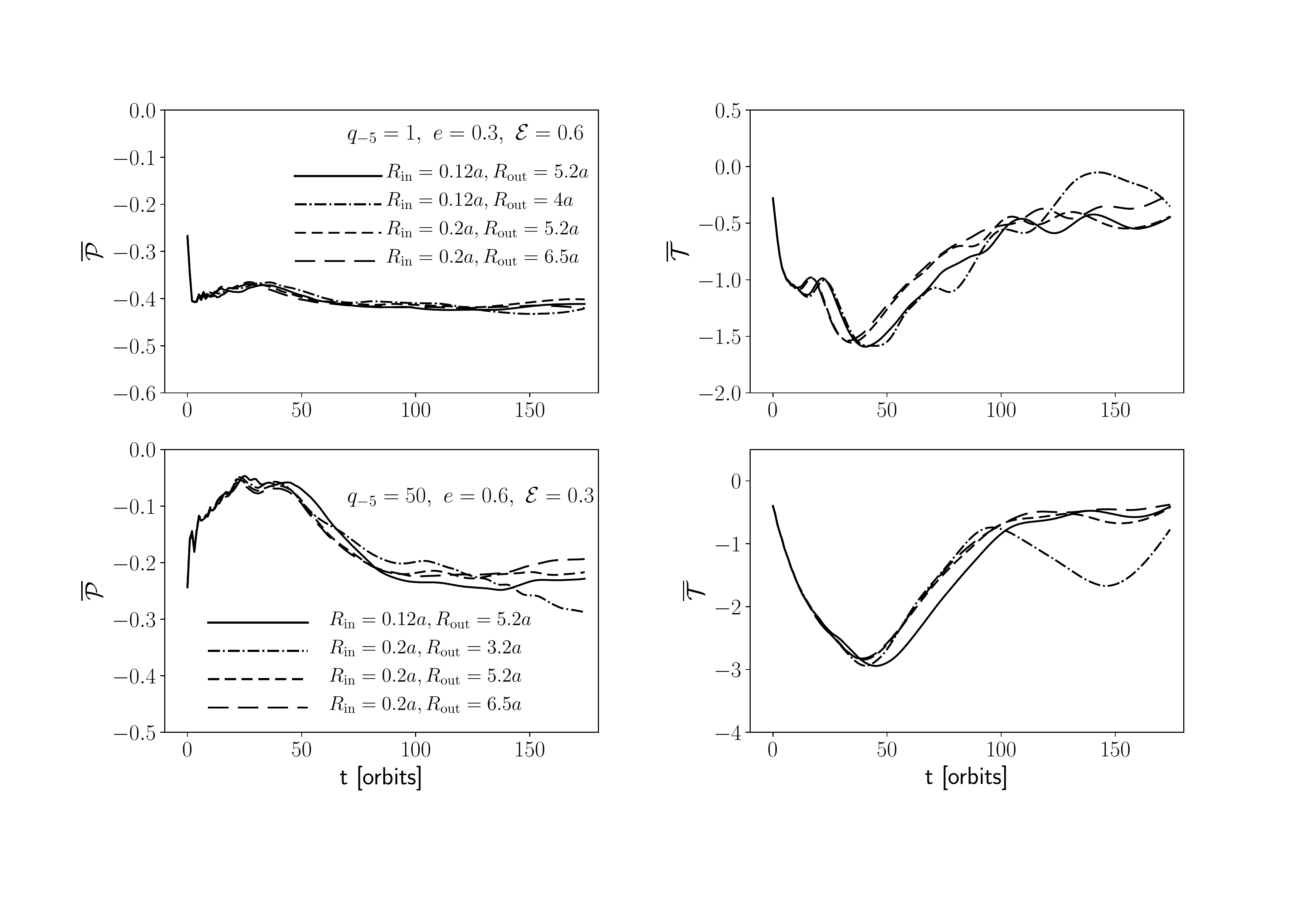}
  \caption{Evolution of the power (left panels) and torque (right panels)
using different radial sizes of the computational box, as quoted in each panel.
 }
\vskip 0.25cm
\label{fig:diff_domains}
\end{figure*}

For completeness, we have computed $\overline{\mathcal{P}}$ and 
$\overline{\mathcal{T}}$ for various combinations of $R_{\rm in}$ and $R_{\rm out}$ 
(see Figure \ref{fig:diff_domains}).
The wave damping rings are $R\in [R_{\rm in},1.3R_{\rm in}]$ and 
$R\in [0.95R_{\rm out},R_{\rm out}]$. We see that $R_{\rm in}=0.2a$ and 
$R_{\rm out}=4a$ are adequate to compute the power within $150$ orbits. 
On the other hand, the torque in runs with $R_{\rm out}\leq 4a$ is not reliable beyond $120$ orbits. Again, it is worth noting that the torque in the simulations
with $R_{\rm in}=0.12a$ and $R_{\rm out}=5.2a$ has the same shape as the torque in the
simulation with $R_{\rm in}=0.2a$ and $R_{\rm out}=5.2a$, but they present a slight shift 
in time.

\section{B. Gradual growth of the mass of the perturber}
\label{sec:adiab_perturber}

\begin{figure*}
\includegraphics[width=199mm,height=85mm]{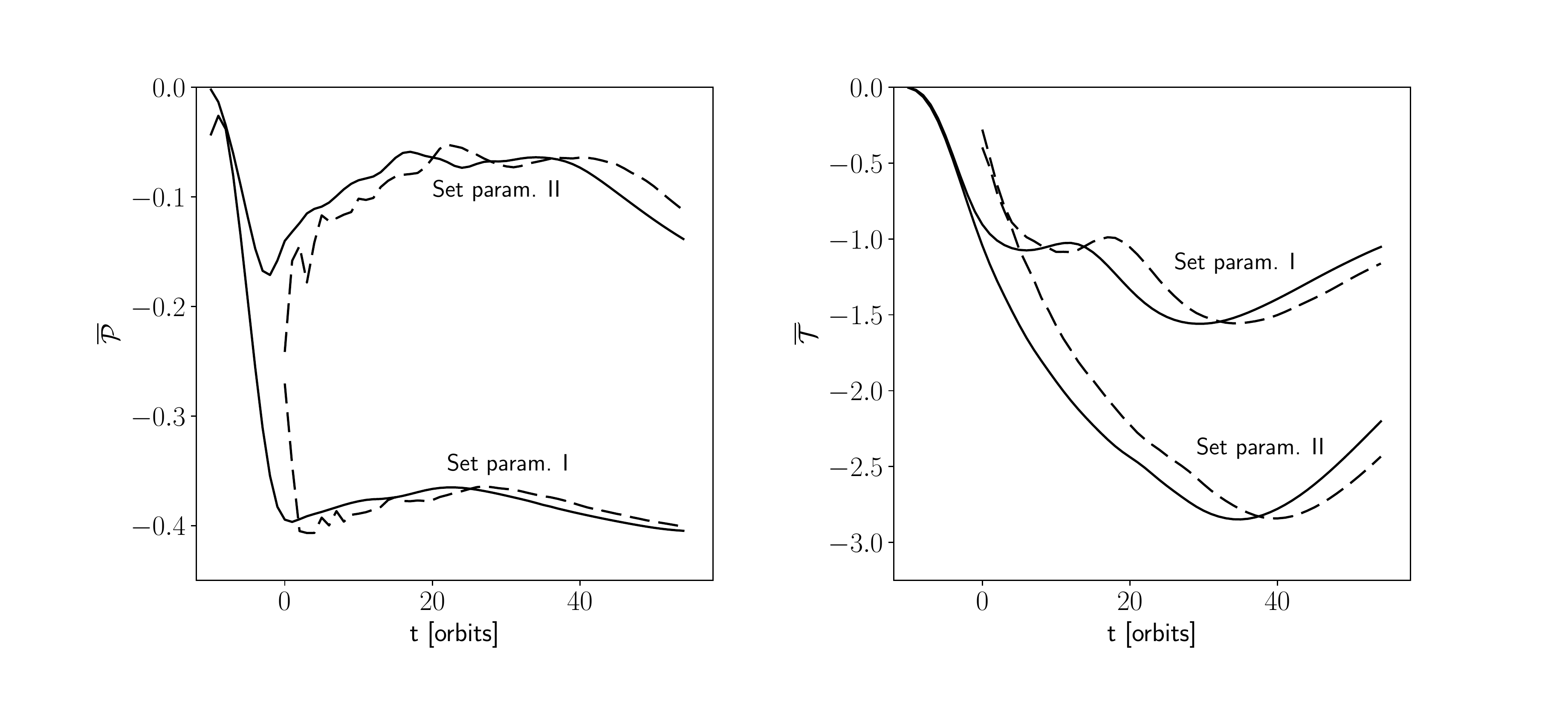}
  \caption{Power (left panel) and torque (right panel) 
when the mass of the perturber increases gradually from 
$t=-10t_{\rm orb}$ to $t=0$ (solid lines) and when the perturber is inserted
suddenly at $t=0$ (dashed lines). The parameters are: $\qmfive=1$, $e=0.3$, 
${\mathcal{E}}=0.6$ (set I)
and $\qmfive=50$, $e=0.6$, ${\mathcal{E}}=0.3$ (set II). In all cases $R_{\rm in}=0.2a$ and
$R_{\rm out}=5.2a$.}
\vskip 0.25cm
\label{fig:mass_taper_1}
\end{figure*}

\begin{figure}
\hskip -0.3cm
\includegraphics[width=92mm,height=76mm]{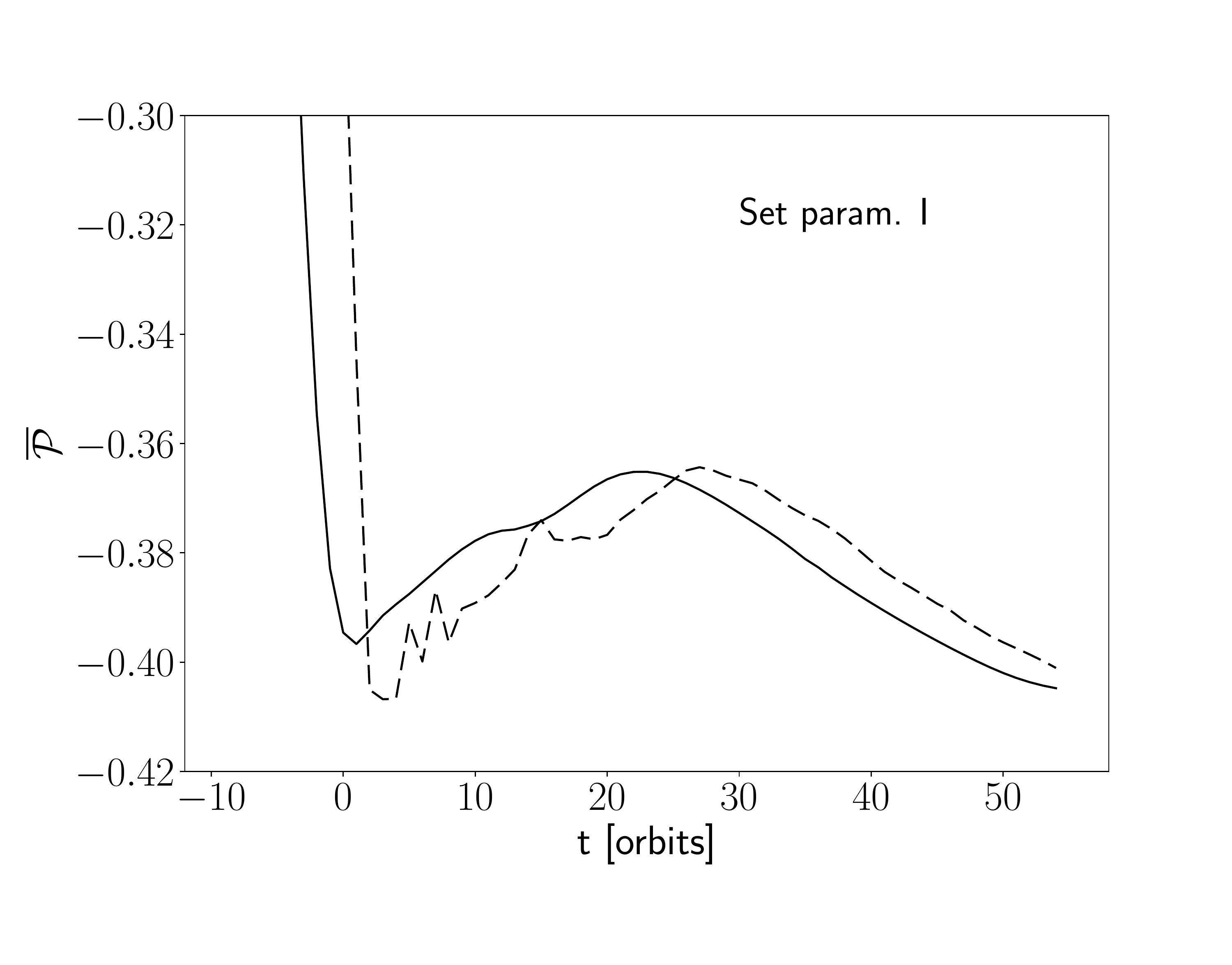}
  \caption{Close-up of the left panel of Figure \ref{fig:mass_taper_1} showing 
the reduction of the wiggles in the power
for the set of parameters I, when the perturber is introduced gradually.
 }
\vskip 0.25cm
\label{fig:pw_mass_taper}
\end{figure}

In models with ${\mathcal{E}}=0.6$, the power exhibits small sawtooth variations during the 
first $20$ orbits (see Figures \ref{fig:diff_ecc}, \ref{fig:diff_rin_stockh} and \ref{fig:diff_domains}).
For retrograde perturbers that are introduced instantaneously, \citet{san18}
showed that these variations in power occur when perturbers catch their own wake.
If so, these transient features should be partially suppressed if the mass of
the perturber gradually increases from $0$ to $M_{p}$ during a time $t_{\scriptscriptstyle M}$ larger 
than $t_{\rm orb}$. Figure \ref{fig:mass_taper_1} shows the power and the torque in simulations where 
we ramp up the mass of the perturber from $0$ at $t=-10t_{\rm orb}$, to $M_{p}$ at $t=0$, 
so that $t_{\scriptscriptstyle M}=10t_{\rm orb}$. For comparison, the results for simulations
in which the perturber is introduced suddenly at $t=0$, so that $t_{\scriptscriptstyle M}=0$,
are also shown. The power and the torque in simulations with 
$t_{\scriptscriptstyle M}=10t_{\rm orb}$ are shifted in time relative to those in 
simulations with $t_{\scriptscriptstyle M}=0$, because at $t=0$
the perturbers have the same mass in both cases, but the perturber with 
$t_{\scriptscriptstyle M}=10t_{\rm orb}$ has
been perturbing the disk during ten orbital periods. As expected, the small wiggles
in the power during the first $15$ orbits 
are suppressed when the perturber is introduced smoothly
(see Figure \ref{fig:pw_mass_taper} for a zoomed-in view).

\end{document}